\documentclass[journal]{IEEEtran}
\usepackage{selectp}

\usepackage{cite}

\usepackage{amsmath,amssymb,amsfonts}
\usepackage{graphicx}
\usepackage{textcomp}
\usepackage{xcolor}
\usepackage{enumitem}
\usepackage{algorithm2e}
\usepackage{verbatim}
\usepackage{amsthm}
\newtheorem{theorem}{Theorem}
\newtheorem{lemma}{Lemma}

\usepackage{url}
\usepackage{bm}
\usepackage[nocomma]{optidef}
\newcommand{\thing}{appendix_on}
\usepackage{ifthen}
\usepackage{hyperref}
\usepackage{cleveref} 
\crefname{lemma}{Lemma}{Lemmas}
\usepackage{ifpdf}
\usepackage{subcaption}
\usepackage{derivative}
\usepackage{dsfont}

\ifCLASSINFOpdf
\else
\fi

\newif\ifincludeComparisonOptimalStaticvsNoSharing

\ifthenelse{\equal{\thing}{appendix_on}}
  {
   \newcommand{\result}{
\begin{appendices}

\section{Proof of Lemma \ref{approx lemma}}
\label{proof_approx}

We use steps similar to those in proof of classic Markov's inequality. For constant $\alpha \geq 0$, we have
\begin{align}
 &  Q_{min} - Q^{i}_{r,n}(k) \geq 0 \implies \left(Q_{min}-Q^{i}_{r,n}(k) + \alpha\right)_+ \geq \alpha, \nonumber \\ & \hspace{2.6cm} \forall \  {n} \in {\mathcal N}^{i}_{r}, \  \forall \  {i} \in {\mathcal O},  \ \forall \ {r} \in {\mathcal R}, \  \forall \  {k} \in \mathcal{K}_n. \nonumber \label{qmin minus q}
\end{align}
Here, the operator $x_+=\max(x,0)$.
Thus,
\begin{align}
    \nonumber 1-\mathds{1} \left\{ Q_{min} - Q^{i}_{r,n}(k)  \leq 0\right\} \leq \frac{\left(Q_{min}-Q^{i}_{r,n}(k) + \alpha\right)_+ }{\alpha},
\end{align}
where $\mathds{1}(.)$ is indicator function.
Taking expectation on both sides of above inequality and bounding its right hand side by 0.05, 
we can show that Lemma \ref{approx lemma} holds.

\ifincludeComparisonOptimalStaticvsNoSharing

\section{Proof of Lemma \ref{comp OSS and NS}}
\label{sectionappendix OSS NS} 
Let ${\mathcal N}^{i}_{r}(k)$ denote the subset of ${\mathcal N}^{i}_{r}$ comprising of clients with packet arrival in period $k$. 

Since all the clients have the same quality scaling factor $\gamma_Q$ and channel capacity $c$, 
Using the model for $Q^{i}_{r,n}$ in \eqref{simqualitymodel}, we  have
\begin{align}
F(\bm{\tau(S),A})  &  = \frac{1}{K} \sum_{k=1}^{K} \sum_{r \in \mathcal {R}} \sum_{i\in \mathcal O} \sum_{{n} \in {\mathcal N}^{i}_{r}(k)} \frac{1}{\gamma_Q} \nonumber \\
& \hspace{2cm}\ln \left(\frac{\frac{\tau_{r,n}^{i}(\bm{S}, k)c }{T}+\theta}{\beta}\right)
 \end{align}
where $\tau^{*,i}_{r,n}(\bm{S})$ is the optimal number of timeslots allotted to the $n^{th}$ client in the $r^{th}$ region of the $i^{th}$ operator if we fix sharing as $\bm{S}$.

Given the objective function is concave and homogeneity of clients, optimal number of timeslots involves equal sharing of available timeslots. Thus, 
\begin{align}
 \tau^{*,i}_{r,n} (\bm{S^*_{NO}})   =   
 \frac{ T }{\sum_{{n} \in {\mathcal N}^{i}_{r}}A_{r,n}^{i}(k)}.
 \end{align}
With sharing, all timeslots in a region are effectively available to all clients (of all operators) of the region and are split equally. Thus,
\begin{align} 
\tau^{*,i}_{r,n} (\bm{S^*_{OSS'}})   =  \frac{
\sum_{j\in {\mathcal O}}S^{*,j\rightarrow i}_{OSS',r}
}{\sum_{i\in \mathcal O}\sum_{{n} \in {\mathcal N}^{i}_{r}}A_{r,n}^{i}(k)} .
 \end{align}

@vinay: switch numeratior @
@vinay: check small o beta term@

\begin{align}
& F(\bm{\tau^*(S^*_{OSS'}),A})-F(\bm{\tau^*(S_{NO}),A}) \nonumber \\
= &\frac{1}{K} \sum_{k=1}^{K} \sum_{r \in \mathcal {R}} \sum_{i\in \mathcal O}  \frac{1}{\gamma_Q} \sum_{{n} \in {\mathcal N}^{i}_{r}}A_{r,n}^{i}(k) \nonumber \\ & \hspace{1cm}\ln \left(\frac{ T \sum_{i\in \mathcal O}\sum_{{n} \in {\mathcal N}^{i}_{r}}A_{r,n}^{i}(k)}{ 
\sum_{j\in {\mathcal O}}S^{*,j\rightarrow i}_{OSS',r}
\sum_{{n} \in {\mathcal N}^{i}_{r}}A_{r,n}^{i}(k)}\right) + o \left(\frac{\beta}{c}\right)
 \end{align}

\fi

\section{Proof of Theorem \ref{theorem_feasibility_optimal}}
\label{feasibilibility_optimal_proof}

Following result regarding $ \textbf{g}(\bm{S}_t, \bm{A}_{t})$ is useful in the proof of the thoerem. Recall (see \eqref{g_definition}) that  $g^{j\rightarrow i}_{r}(\mathbf{S}_t, \mathbf{A}_{t})=-\sum_{k \in \mathcal{P}_t}{\lambda}_{r}^{i} 
  (\bm{S}_t,\bm{A}(k))$.

\begin{lemma}
\label{gradient in terms of lagrange multipliers}
For any hyperperiod $t$,
\begin{align}
   \textbf{g}(\bm{S}_t, \bm{A}_{t}) = \nabla_{\bm{S}_t} \left( \sum_{k \in \mathcal{P}_t} F_k(\bm{\tau}^*(\bm{S}_t,k),\bm{A}_{t})\right).
\end{align}

 \end{lemma}
  \begin{proof}
  Recall from  \eqref{g_definition} that $g^{j\rightarrow i}_{r}(\mathbf{S}_t, \mathbf{A}_{t})=-\sum_{k \in \mathcal{P}_t}{\lambda}_{r}^{i} 
  (\bm{S}_t,\bm{A}(k)), \ \forall {i, \ j} \in {\mathcal O}, \forall{r} \in {\mathcal R}$.
  Lemma \ref{gradient in terms of lagrange multipliers} concerns sensitivity of optimal value of objective $F_k(\bm{\tau}^*(\bm{S}_t,k),\bm{A}_{t}) $ of optimization problem 
 RA$(\bm{S}_t,\bm{A}(k))$ to small perturbations of $\bm{S}_t$.
Lemma \ref{gradient in terms of lagrange multipliers} follows from  results in Section 5.6 of \cite{boyd2004convex} which relates optimal Lagrange multiplier to sensitivity (via partial-derivative) of the optimal value of objective function.
 \end{proof}

 \begin{lemma}\label{optimal_value_of_Fk_is_convex_function}
 $F_k(\bm{\tau}^*(\bm{S}_t,k),\bm{A}_{t})$ is a convex function of $\bm{S}_t$ 
 \end{lemma}
  \begin{proof} 
  We omit a detailed proof. It primarily uses Corollary 2.2 of \cite{fiacco1986convexity}.
 \end{proof}

For any period $k$, let $\Theta^*(k) = \left(\bm{P}^*(k), \bm{S}^*_{t(k)} \right)$ where $t(k)=\lceil{k/H} \rceil$ is the hyperperiod's index associated with period $k$.
Recall that  
$\bm{P}^*(k)$ is the virtual queue evolving as \eqref{update eqn} and $\bm{S}^*_{t(k)}$ evolves according to \eqref{update S}.
Define quadratic Lyapunov function $L(\Theta^*(k))$ as
\begin{align}
     L(\Theta^*(k)) = & \frac{1}{2}  \left [ \sum_{r \in \mathcal {R}} \sum_{i\in \mathcal O}\sum_{{n} \in {\mathcal N}^{i}_{r}}  P_{r,n}^{*i}(k)^2 \right] + \frac{1}{2 \eta}\lVert \bm{S}^*_{t(k)}-\bm{S^{**}} \rVert^2 , \nonumber \\ & \hspace{4cm} \forall { k \in \mathcal K}.\label{lyapunov4}
\end{align}
 where $\bm{S}^{**}$ is the optimal static sharing vector of an asymptotically optimal stationary randomized policy
with static sharing obtained by solving  OPT-SS*$_K(\bm{A})$ as $K \to \infty$.

\begin{lemma}\label{drift- quality lemma3}
 There exist 
 positive constants $B$, $C$ and $D$ such that for each period $k\in\{0,H,2H...\}$, we have
\begin{align}
& \mathbb{E}\left[L(\Theta^*(k+H))-L(\Theta^*(k)) | \Theta^*(k)\right] - \nonumber \\ &  V \mathbb{E} \left[  \sum_{r \in \mathcal {R}} \sum_{i\in \mathcal O}\sum_{{n} \in {\mathcal N}^{i}_{r}} \sum_{k' \in \mathcal{P}_{t(k)}} \hspace{-0.2cm} Q \left(\tau_{r,n}^{*i}(\bm{S}^{*}_{t(k)},k')\right)A^{i}_{r,n}(k')| \Theta^*(k) \right]  \nonumber \\
& \leq  B+D +  \frac{\eta C}{2}  
 - \frac{\epsilon}{2}p  \sum_{r \in \mathcal {R}} \sum_{i\in \mathcal O}\sum_{{n} \in {\mathcal N}^{i}_{r}} \sum_{k' \in \mathcal{P}_{t(k),n}} \hspace{-0.2cm} P_{r,n}^{*i}(k')  - \nonumber \\ &    V \mathbb{E} \left[ \sum_{r \in \mathcal {R}} \sum_{i\in \mathcal O}\sum_{{n} \in {\mathcal N}^{i}_{r}} \sum_{k' \in \mathcal{P}_{t(k)}} \hspace{-0.2cm}  Q \left(\tau_{r,n}^{*i}(\bm{S}^{**},k')\right)A^{i}_{r,n}(k') | \Theta^*(k)\right]  
   \end{align}  
 \end{lemma}

 \begin{proof}

 $t+1$ can be considered equivalent to $k+H$\\

 Let $(Q_{min}-Q(\tau^{*i}_{r,n}(k))+\alpha)_+$ be denoted by $X^{*i}_{r,n}(\bm{S}^*_{t(k)},k)$ and $0.05 \alpha$ by $p$.

Using induction and \eqref{update eqn}, we can show that
\begin{align}
   & P_{r,n}^{*i}(k+H)  \nonumber   \leq \max\left( P_{r,n}^{*i}(k)  - \sum_{k' \in \mathcal{P}_{t(k),n}} p ,0 \right) \nonumber \\ & +\sum_{k' \in \mathcal{P}_{t(k),n}} X^{*i}_{r,n}(\bm{S}^*_{t(k)},k') , \qquad \forall n \in \mathcal{N}^{i}_{r}, \ i\in \mathcal{O}, \ r \in \mathcal{R}
\end{align}

Using Lemma 4.3 from \cite{neely},
\begin{align}
    & P_{r,n}^{*i}(k+H)^2 - P_{r,n}^{*i}(k)^2 \nonumber \\ & \leq   \left(\sum_{k' \in \mathcal{P}_{t(k),n}} p \right)^2 + \left(\sum_{k' \in \mathcal{P}_{t(k),n}} X^{*i}_{r,n}(\bm{S}^*_{t(k)},k') \right)^2 \nonumber \\ & + 2 P_{r,n}^{*i}(k) \left(\sum_{k' \in \mathcal{P}_{t(k),n}} \left [X^{*i}_{r,n}(\bm{S}^*_{t(k)},k') - p \right ] \right). \nonumber 
\end{align}
We can modify the above inequality to obtain 
\begin{align}
    & P_{r,n}^{*i}(k+H)^2 - P_{r,n}^{*i}(k)^2  \nonumber \\  & 
    \leq B  + D +  2 \sum_{k' \in \mathcal{P}_{t(k),n}} [P_{r,n}^{*i}(k') ]\left [X^{*i}_{r,n}(\bm{S}^*_{t(k)},k') - p \right ],
\end{align}
where $B$ is an upper bound on (the bounded terms)
\begin{align*}
 B(k) & = \left(\sum_{k' \in \mathcal{P}_{t(k),n}} p \right)^2 + \left(\sum_{k' \in \mathcal{P}_{t(k),n}} X^{*i}_{r,n}(\bm{S}^*_{t(k)},k') \right)^2, 
\end{align*}
and $D$ is an upper bound on (the bounded terms)
\begin{align*} 
    D(k) & = [ P_{r,n}^{*i}(k) - P_{r,n}^{*i}(k') ]\left [X^{*i}_{r,n}(\bm{S}^*_{t(k)},k')- 0.05\alpha \right ].
\end{align*}

\begin{align}
  & L(\Theta^*(k+H))-L(\Theta^*(k))  \nonumber \\ \leq & B + D \nonumber \\ & +  \sum_{r \in \mathcal {R}} \sum_{i\in \mathcal O}\sum_{{n} \in {\mathcal N}^{i}_{r}}\sum_{k' \in \mathcal{P}_{t(k),n}} P_{r,n}^{*i}(k')  \left [X^{*i}_{r,n}(\bm{S}^*_{t(k)},k') - p \right ]   \nonumber \\ & + \frac{1}{2 \eta}\lVert \bm{S}^*_{t(k)+1}-\bm{S^{**}} \rVert^2  - \frac{1}{2 \eta} \lVert \bm{S}^*_{t(k)}-\bm{S^{**}} \rVert^2 
  \label{lyapunov_delta_int_step}
         \end{align}

The update rule of iterative semi-static sharing policy is given by :
\[
\bm{S}_{t(k)+1} = \Pi_{\Omega}(\bm{S}_{t(k)} - \eta \bm{g}(\bm{S}_{t(k)}, \bm{P}_{t(k)},\bm{A}_{t(k)}))
\]

Using property of projection operators, we have
\begin{align}
& \mathbb{E}[\|\bm{S}^*_{t(k)+1} - \bm{S}^{**} \|^2 | \Theta^*(k)]  - \mathbb{E}[\|\bm{S}^*_{t(k)} - \bm{S}^{**}\|^2| \Theta^*(k)] \nonumber \\ & \leq   - 2\eta \mathbb{E}[\langle \bm{g}(\bm{S}^*_{t(k)}, \bm{P}_{t(k)},\bm{A}_{t(k)})), \bm{S}^*_{t(k)} - \bm{S}^{**} \rangle| \Theta^*(k)] \nonumber \\ &  + \eta^2 \mathbb{E}[\|\bm{g}(\bm{S}^*_{t(k)},\bm{P}_{t(k)},\bm{A}_{t(k)}))\|^2| \Theta^*(k)] \nonumber \\
 & = -2\eta \langle \mathbb{E}[ \bm{g}(\bm{S}^*_{t(k)},\bm{P}_{t(k)},\bm{A}_{t(k)})) | \Theta^*(k)],(\bm{S}^*_{t(k)} - \bm{S}^{**}) \rangle  \nonumber \\ & + \eta^2 \mathbb{E}[\|\bm{g}(\bm{S}^*_{t(k)},\bm{P}_{t(k)},\bm{A}_{t(k)}))\|^2| \Theta^*(k)] 
 \label{projection_operator_property_last_step}
\end{align}

Applying conditional expectation on Lemma \ref{gradient in terms of lagrange multipliers}, we have
\begin{align}
   & \mathbb{E}[\bm{g}(\bm{S}^*_{t(k)}, \bm{P}_{t(k)},\bm{A}_{t(k)})) \mid  \Theta^*(k)] \nonumber \\ & = \mathbb{E} \left[ \sum_{k \in \mathcal{P}_{t(k),n}} \nabla_{\bm{S}^*_{t(k)}}  F_k(\bm{\tau}^*(\bm{S}^*_{t(k)}),\bm{P}(k),\bm{A}(k)))\mid  \Theta^*(k) \right] \nonumber 
\end{align}
Let 
$$f(\bm{S}^*_{t(k)})= \mathbb{E} \left[ \sum_{k \in \mathcal{P}_{t(k),n}} F_k(\bm{\tau^*}(\bm{S}^*_{t(k)}), \bm{P}^*(k), \bm{A}(k))| \Theta^*(k) \right].$$
Note that gradient of $F_k(\bm{\tau}^*(.),\bm{P}(k),\bm{A}(k)))$ above is bounded as $Q$ is twice-differentiable on a compact set.
Applying Dominated Convergence Theorem (Theorem 11.12, \cite{hajek2015random}),
\begin{align}
  \mathbb{E}[\bm{g}(\bm{S}^*_{t(k)}, \bm{P}_{t(k)}^*,\bm{A}_{t(k)})) \mid  \Theta^*(k)] = \nabla_{\bm{S}^*_{t(k)}} f(\bm{S}^*_{t(k)}) \label{unbiased_estimator3}  
\end{align}

Using the above result in \eqref{projection_operator_property_last_step},
we have
\begin{align}
 & \mathbb{E}[\|\bm{S}^*_{t(k)+1} - \bm{S}^{**} \|^2 | \Theta^*(k)]  - \mathbb{E}[\|\bm{S}^*_{t(k)} - \bm{S}^{**}\|^2| \Theta^*(k)]  \nonumber \\ & \leq - 2\eta \langle \nabla_{\bm{S}^*_{t(k)}} f(\bm{S}^*_{t(k)}), \bm{S}^*_{t(k)} - \bm{S}^{**} \rangle \nonumber \\ &  + \eta^2 \mathbb{E}[\|\bm{g}(\bm{S}^*_{t(k)}, \bm{P}_{t(k)}^*,\bm{A}_{t(k)}))\|^2| \Theta^*(k)]. \label{simplified with inner prdt remaining2} 
 \\
&\leq - 2\eta [f(\bm{S}^*_{t(k)}) - f(\bm{S}^{**})]  \nonumber \\ & + \eta^2 \mathbb{E}[\|\bm{g}(\bm{S}^*_{t(k)}, \bm{P}_{t(k)}^*,\bm{A}_{t(k)}))\|^2| \Theta^*(k)]. \label{after using f con3}
 \\
&\leq - 2\eta 
 \mathbb{E} \left[ \sum_{k \in \mathcal{P}_{t(k),n}} F_k(\bm{\tau^*}(\bm{S}^*_{t(k)}), \bm{P}^*(k), \bm{A}(k))| \Theta^*(k) \right]   \nonumber
 \\& + 2\eta  \mathbb{E} \left[ \sum_{k \in \mathcal{P}_{t(k),n}} F_k(\bm{\tau^*}(\bm{S}^{**}), \bm{P}^*(k), \bm{A}(k))| \Theta^*(k) \right]
 \nonumber 
\\ & + \eta^2 \mathbb{E}[\|\bm{g}(\bm{S}^*_{t(k)}, \bm{P}_{t(k)}^*,\bm{A}_{t(k)}))\|^2| \Theta^*(k)]. \label{expand using f definition}
\end{align}
where we have used convexity of $f$ to obtain the inequality in \eqref{after using f con3}. The convexity of $f$ follows from Lemma \ref{optimal_value_of_Fk_is_convex_function}.
Let $C$ denote an upper bound on the bounded function $\|\bm{g}(\bm{S}^*_{t(k)}, \bm{P}_{t(k)}^*,\bm{A}_{t(k)}))\|^2 $.

Using \eqref{expand using f definition} in \eqref{lyapunov_delta_int_step} and expanding $F_k$, we get
\begin{align}
& \mathbb{E}\left[L(\Theta^*(k+H))-L(\Theta^*(k)) | \Theta^*(k)\right] \nonumber \\
& \leq B+D+  \frac{\eta C}{2} + \nonumber \\ & 
  \sum_{r \in \mathcal {R}} \sum_{i\in \mathcal O}\sum_{{n} \in {\mathcal N}^{i}_{r}} \hspace{-0.1cm} \mathbb{E} \hspace{-0.1cm}  \left[\sum_{k' \in \mathcal{P}_{t(k),n}} \hspace{-0.3cm}P_{r,n}^{*i}(k') \left [X^{*i}_{r,n}(\bm{S}^*_{t(k)},k') - p \right ] | \Theta^*(k) \right ]\nonumber \\ 
& + V \mathbb{E} \hspace{-0.1cm} \left[  \sum_{r \in \mathcal {R}} \sum_{i\in \mathcal O}\sum_{{n} \in {\mathcal N}^{i}_{r}} \sum_{k' \in \mathcal{P}_{t(k)}} \hspace{-0.3cm}  Q \left(\tau_{r,n}^{*i}(\bm{S}^{*}_{t(k)},k')\right)A^{i}_{r,n}(k')| \Theta^*(k) \right] \nonumber \\ & -  V \mathbb{E} \left[ \sum_{r \in \mathcal {R}} \sum_{i\in \mathcal O}\sum_{{n} \in {\mathcal N}^{i}_{r}} \sum_{k' \in \mathcal{P}_{t(k)}} \hspace{-0.3cm} Q \left(\tau_{r,n}^{*i}(\bm{S}^{**},k')\right)A^{i}_{r,n}(k') | \Theta^*(k)\right] \nonumber \\ & -  \mathbb{E} \left[\sum_{r \in \mathcal {R}} \sum_{i\in \mathcal O} \sum_{{n} \in {\mathcal N}^{i}_{r}} \sum_{k' \in \mathcal{P}_{t(k),n}} P_{r,n}^{*i}(k') X^{*i}_{r,n}(\bm{S}^*_{t(k)},k') | \Theta^*(k) \right] \nonumber \\
& +  \mathbb{E} \left[\sum_{r \in \mathcal {R}} \sum_{i\in \mathcal O} \sum_{{n} \in {\mathcal N}^{i}_{r}} \sum_{k' \in \mathcal{P}_{t(k),n}} P_{r,n}^{*i}(k') X^{*i}_{r,n}(\bm{S}^{**},k') | \Theta^*(k)\right] 
   \end{align}

Using optimality of $\tau_{r,n}^{*i}(\bm{S}^{**},k')$ for 
RA$(\bm{S}^{**},\bm{A}(k'))$, we get the following inequality on replacing 
$\tau_{r,n}^{*i}(\bm{S}^{**},k')$ 
with
$\tau_{r,n}^{**i}(\bm{S}^{**},k')$
in the above inequality,
\begin{align} 
& \mathbb{E}\left[L(\Theta^*(k+H))-L(\Theta^*(k)) | \Theta^*(k)\right] \nonumber \\
& \leq B+ D +  \frac{\eta C}{2}  
 - p  \sum_{r \in \mathcal {R}} \sum_{i\in \mathcal O}\sum_{{n} \in {\mathcal N}^{i}_{r}} \sum_{k' \in \mathcal{P}_{t(k),n}} P_{r,n}^{*i}(k') + \nonumber \\ 
&  V \mathbb{E} \left[  \sum_{r \in \mathcal {R}} \sum_{i\in \mathcal O}\sum_{{n} \in {\mathcal N}^{i}_{r}} \sum_{k' \in \mathcal{P}_{t(k)}} \hspace{-0.3cm} Q \left(\tau_{r,n}^{*i}(\bm{S}^{*}_{t(k)},k')\right)A^{i}_{r,n}(k')| \Theta^*(k) \right] \nonumber 
\\ & -  V \mathbb{E} \left[ \sum_{r \in \mathcal {R}} \sum_{i\in \mathcal O}\sum_{{n} \in {\mathcal N}^{i}_{r}} \sum_{k \in \mathcal{P}_{t(k)}}   \hspace{-0.3cm}Q \left(\tau_{r,n}^{**i}(\bm{S}^{**},k')\right)A^{i}_{r,n}(k') | \Theta^*(k)\right] \nonumber \\ 
& +  \mathbb{E} \left[\sum_{r \in \mathcal {R}} \sum_{i\in \mathcal O} \sum_{{n} \in {\mathcal N}^{i}_{r}} \sum_{k' \in \mathcal{P}_{t(k),n}} P_{r,n}^{*i}(k') X^{*i}_{r,n}(\bm{S}^{**},k') | \Theta^*(k)\right] 
   \end{align}
Rearranging the terms, we get
 \begin{align}
& \mathbb{E}\left[L(\Theta^*(k+H))-L(\Theta^*(k)) | \Theta^*(k)\right] - \nonumber \\ &   V \mathbb{E} \left[  \sum_{r \in \mathcal {R}} \sum_{i\in \mathcal O}\sum_{{n} \in {\mathcal N}^{i}_{r}} \sum_{k' \in \mathcal{P}_{t(k)}} \hspace{-0.3cm} Q \left(\tau_{r,n}^{*i}(\bm{S}^{*}_{t(k)},k')\right)A^{i}_{r,n}(k')| \Theta^*(k) \right]  \nonumber \\
& \leq B + D +  \frac{\eta C}{2}  
 - p  \sum_{r \in \mathcal {R}} \sum_{i\in \mathcal O}\sum_{{n} \in {\mathcal N}^{i}_{r}} \sum_{k' \in \mathcal{P}_{t(k),n}} P_{r,n}^{*i}(k') - \nonumber \\ 
&   V \mathbb{E} \left[ \sum_{r \in \mathcal {R}} \sum_{i\in \mathcal O}\sum_{{n} \in {\mathcal N}^{i}_{r}} \sum_{k' \in \mathcal{P}_{t(k)}}  \hspace{-0.3cm} Q \left(\tau_{r,n}^{**i}(\bm{S}^{**},k')\right)A^{i}_{r,n}(k') | \Theta^*(k)\right] \nonumber \\ 
& +  \mathbb{E} \left[\sum_{r \in \mathcal {R}} \sum_{i\in \mathcal O} \sum_{{n} \in {\mathcal N}^{i}_{r}} \sum_{k' \in \mathcal{P}_{t(k),n}} P_{r,n}^{*i}(k') X^{*i}_{r,n}(\bm{S}^{**},k') | \Theta^*(k)\right] 
   \end{align}  

 Since $Q_{min}$ is strictly static feasible, there exists a constant $\kappa$ with $0<\kappa<1$, and  stationary randomized policies
with static sharing that satisfy \eqref{nonneg}, \eqref{maxshare}, \eqref{maxschedule}, \eqref{sharing constraint_policy}, \eqref{modified_convexified_constraint}. Among these policies, there exists an asymptotically optimal stationary randomized policy
with static sharing taking optimal static sharing $\bm{S}^{**}$ and client resource allocation $\tau_{r,n}^{**i}(\bm{S}^{**},k)$ in period $k$. Thus, we have
\begin{align}
 & E\left [X^{*i}_{r,n}(\bm{S}^{**},k')  \middle |\Theta^*(k)\right ]  \leq \left (1-\frac{\epsilon}{2} \right ) p,  \nonumber \\ &  \hspace{2cm} \forall n\in {\mathcal N}^{i}_{r}(k),\  \forall{r} \in {\mathcal R}, \  \forall {i} \in {\mathcal O} \label{mu simp4}
 \end{align}
where $\epsilon=1-\kappa$.

Thus, we have
\begin{align}
& E\left[L(\Theta^*(k+H))-L(\Theta^*(k)) | \Theta^*(k)\right] \nonumber \\ & -  V \mathbb{E} \left[  \sum_{r \in \mathcal {R}} \sum_{i\in \mathcal O}\sum_{{n} \in {\mathcal N}^{i}_{r}} \sum_{k' \in \mathcal{P}_{t(k)}}  Q \left(\tau_{r,n}^{*i}(\bm{S}^{*}_{t(k)},k')\right)A^{i}_{r,n}(k')| \Theta^*(k) \right]  \nonumber \\
& \leq B+D+  \frac{\eta C}{2}  
 - p  \sum_{r \in \mathcal {R}} \sum_{i\in \mathcal O}\sum_{{n} \in {\mathcal N}^{i}_{r}} \sum_{k' \in \mathcal{P}_{t(k),n}} P_{r,n}^{*i}(k') \nonumber \\ 
& -  V \mathbb{E} \left[ \sum_{r \in \mathcal {R}} \sum_{i\in \mathcal O}\sum_{{n} \in {\mathcal N}^{i}_{r}} \sum_{k' \in \mathcal{P}_{t(k)}}  Q \left(\tau_{r,n}^{*i}(\bm{S}^{**},k')\right)A^{i}_{r,n}(k') | \Theta^*(k)\right] \nonumber \\ & + \sum_{r \in \mathcal {R}} \sum_{i\in \mathcal O} \sum_{{n} \in {\mathcal N}^{i}_{r}} \sum_{k' \in \mathcal{P}_{t(k),n}} P_{r,n}^{*i}(k') \left (1-\frac{\epsilon}{2} \right ) p \\
& = B + D +  \frac{\eta C}{2}  
 - \frac{\epsilon}{2}p  \sum_{r \in \mathcal {R}} \sum_{i\in \mathcal O}\sum_{{n} \in {\mathcal N}^{i}_{r}} \sum_{k' \in \mathcal{P}_{t(k),n}} P_{r,n}^{*i}(k') \nonumber \\ &  -  V \mathbb{E} \left[ \sum_{r \in \mathcal {R}} \sum_{i\in \mathcal O}\sum_{{n} \in {\mathcal N}^{i}_{r}} \sum_{k' \in \mathcal{P}_{t(k)}}  Q \left(\tau_{r,n}^{*i}(\bm{S}^{**},k')\right)A^{i}_{r,n}(k') | \Theta^*(k)\right] \nonumber 
   \end{align}  
\end{proof}

\begin{lemma}\label{average queue lemma3}
There exist 
 positive constants $B,  C$ and $D$ such that for any positive integer $M$
\begin{align}
 & \limsup_{M \rightarrow \infty} \frac{1}{M}\sum_{k=0}^{MH-1}\mathbb{E} \left[\frac{\epsilon}{2} p  \sum_{r \in \mathcal {R}} \sum_{i\in \mathcal O}\sum_{{n} \in {\mathcal N}^{i}_{r}}  P_{r,n}^{*i}(k) A^{i}_{r,n}(k) \right]  \nonumber \\ & \leq  B + D  + \frac{\eta C}{2} -\frac{E\left [ {L}(\Theta^*(MH))\right ]}{M}   \nonumber 
 \\
 &  + \limsup_{M \rightarrow \infty}\frac{V}{M}\sum_{k=0}^{MH-1}  E \left [\sum_{r \in \mathcal {R}} \sum_{i\in \mathcal O}\sum_{{n} \in {\mathcal N}^{i}_{r}} Q\left(\tau_{r,n}^{*i}(\bm{S}^*_{t(k)},k)\right)A^{i}_{r,n}(k) \right] \nonumber 
 \\ & -  \limsup_{M \rightarrow \infty}\frac{V}{M}\sum_{k=0}^{MH-1}   E \left [\sum_{r \in \mathcal {R}} \sum_{i\in \mathcal O}\sum_{{n} \in {\mathcal N}^{i}_{r}}   Q\left(\tau_{r,n}^{**i}(\bm{S}^{**},\bm{A})\right)A^{i}_{r,n}(k) \right]  
 \end{align}
 \end{lemma}

  \begin{proof}
Taking expectation on both sides of the inequality in Lemma \ref{drift- quality lemma3} statement, and applying the law of iterated expectations, we have
\begin{align}
 &  E\left [ {L}(\Theta^*(k+H)\right ]-E\left [{L}(\Theta^*(k))\right ] \nonumber \\ & -V \mathbb{E} \left [\sum_{r \in \mathcal {R}} \sum_{i\in \mathcal O}\sum_{{n} \in {\mathcal N}^{i}_{r}} \sum_{k' \in \mathcal{P}_{t(k)}} Q\left(\tau_{r,n}^{*i}(\bm{S}^*_{t(k)},k')\right)A^{i}_{r,n}(k') \right] \nonumber 
 \\ & \leq B+D + \frac{\eta C}{2} \nonumber \\ & - V \mathbb{E} \left [ \sum_{r \in \mathcal {R}} \sum_{i\in \mathcal O}\sum_{{n} \in {\mathcal N}^{i}_{r}} \sum_{k' \in \mathcal{P}_{t(k)}} Q\left(\tau_{r,n}^{**i}(\bm{S}^{**},\bm{A})\right)A^{i}_{r,n}(k')\right] \nonumber \\ & - \mathbb{E} \left[\frac{\epsilon}{2} p \sum_{r \in \mathcal {R}} \sum_{i\in \mathcal O}\sum_{{n} \in {\mathcal N}^{i}_{r}} \sum_{k' \in \mathcal{P}_{t(k)}} P_{r,n}^{*i}(k')A^{i}_{r,n}(k') \right]\label{iterated3}
 \end{align}

 Summing above inequality over periods $k$ from $0, H, 2H, ..., (M-1)H$, cancelling terms in the resulting telescoping series, and combining the summations over periods $k$ from $0, H, 2H, ..., (M-1)H$ and $ \sum_{k' \in \mathcal{P}_{t(k)}}$, we have 
\begin{align}
 &  E\left [ {L}(\Theta^*(MH))\right ]-E\left [{L}(\Theta^*(0))\right ] \nonumber 
 \\ &  - V \sum_{k=0}^{MH-1}  E \left [\sum_{r \in \mathcal {R}} \sum_{i\in \mathcal O}\sum_{{n} \in {\mathcal N}^{i}_{r}} Q\left(\tau_{r,n}^{*i}(\bm{S}^*_{t(k)},k)\right)A^{i}_{r,n}(k) \right] \nonumber \\ & \leq BM + DM   + \frac{\eta C M}{2}  \nonumber 
 \\ & - V  \sum_{k=0}^{MH-1} E \left [\sum_{r \in \mathcal {R}} \sum_{i\in \mathcal O}\sum_{{n} \in {\mathcal N}^{i}_{r}}  
 Q\left(\tau_{r,n}^{**i}(\bm{S}^{**},\bm{A})\right)A^{i}_{r,n}(k)\right] \nonumber 
 \\ & - \sum_{k=0}^{MH-1} \mathbb{E} \left[\frac{\epsilon}{2} p \sum_{r \in \mathcal {R}} \sum_{i\in \mathcal O}\sum_{{n} \in {\mathcal N}^{i}_{r}}  P_{r,n}^{*i}(k) A^{i}_{r,n}(k)\right]
 \end{align}
 
Dividing above inequality by $M$, using the fact that $ {L(\Theta^*(0))} \geq 0$ and rearranging the terms, we have
\begin{align}
 & \frac{1}{M} \sum_{k=0}^{MH-1} \mathbb{E} \left[\frac{\epsilon}{2} p  \sum_{r \in \mathcal {R}} \sum_{i\in \mathcal O}\sum_{{n} \in {\mathcal N}^{i}_{r}} P_{r,n}^{*i}(k) A^{i}_{r,n}(k) \right]  \nonumber 
 \\ & \leq  B + D + \frac{\eta C}{2} -\frac{E\left [ {L}(\Theta^*(MH))\right ]}{M}   \nonumber \\
 &  + \frac{V}{M}\sum_{k=0}^{MH-1}  E \left [\sum_{r \in \mathcal {R}} \sum_{i\in \mathcal O}\sum_{{n} \in {\mathcal N}^{i}_{r}} Q\left(\tau_{r,n}^{*i}(\bm{S}^*_{t(k)},k)\right)A^{i}_{r,n}(k) \right] \nonumber 
 \\ & - \frac{V}{M}\sum_{k=0}^{MH-1} E \left [\sum_{r \in \mathcal {R}} \sum_{i\in \mathcal O}\sum_{{n} \in {\mathcal N}^{i}_{r}}   Q\left(\tau_{r,n}^{**i}(\bm{S}^{**},\bm{A})\right)A^{i}_{r,n}(k) \right]  \label{before limsup3}
 \end{align}

Lemma \eqref{average queue lemma3} follows by taking $\limsup_{M \rightarrow \infty}$ of the above inequality.

\end{proof}

 \begin{lemma}\label{average quality lemma3}
There exists 
 positive constants $B,  C$ and $D$ such that
\begin{align}
 &  \liminf_{M \to \infty} \frac{1}{M} \sum_{r \in \mathcal {R}} \sum_{i\in \mathcal O}\sum_{{n} \in {\mathcal N}^{i}_{r}} \sum_{k=0}^{MH-1}  E \left[Q(\tau_{r,n}^{*i}(\bm{S}^*_{t(k)},k))A^{i}_{r,n}(k)\right]    \nonumber 
 \\ & \geq  E \left [\sum_{r \in \mathcal {R}} \sum_{i\in \mathcal O}\sum_{{n} \in {\mathcal N}^{i}_{r}} \sum_{k' \in \mathcal{P}_{t(k)}} Q\left(\tau_{r,n}^{**i}(\bm{S}^{**},\bm{A})\right)A^{i}_{r,n}(k')\right] \nonumber \\ & -\frac{B+D}{V} - \frac{\eta C}{2 V}
 \end{align}

 \end{lemma}

\begin{proof}

 Dividing \eqref{before limsup3} by $V$ and noting that $ {L(\Theta^*(MH))} \geq 0$ and $P_{r,n}^{*i}(k)\geq 0$, we have

 \begin{align}
 & \frac{1}{M} \sum_{r \in \mathcal {R}} \sum_{i\in \mathcal O}\sum_{{n} \in {\mathcal N}^{i}_{r}} \sum_{k=0}^{MH-1}  E \left[Q(\tau_{r,n}^{*i}(\bm{S}^*_{t(k)},k))A^{i}_{r,n}(k)\right]  \nonumber \\ 
 & \geq E \left [\sum_{r \in \mathcal {R}} \sum_{i\in \mathcal O}\sum_{{n} \in {\mathcal N}^{i}_{r}} \sum_{k' \in \mathcal{P}_{t(k)}} Q\left(\tau_{r,n}^{**i}(\bm{S}^{**},\bm{A})\right)A^{i}_{r,n}(k') \right] \nonumber \\ & -\frac{B+D}{V} - \frac{\eta C}{2 V}
 \end{align}
 
Taking $\liminf_{M \to \infty}$ on both sides of the above inequality, the result is obtained.
\end{proof}

 Following result is from \cite{prk} given below for easy reference.
\begin{lemma} \textbf{(from \cite{prk})} \label{prk_lemma1}
 Let $f(n)$ be a nonnegative function such that $|f(n+1)-f(n)| \leq M$, for some $M>0$, for all n. \\
 If $\limsup_{n \to \infty}\frac{1}{n}\sum_{i=0}^n f(i) \leq B$, for some constant B, then $\lim_{n \to \infty}\frac{1}{n}f(n)=0$ 
\end{lemma}

\begin{lemma}\label{lemma_new0}
For any strictly static feasible $Q_{min}$,  we have
 \begin{align}
 & \lim_{M \to \infty} \frac{1}{M}  E \left[  P_{r,n}^{*i}(MH)A_{r,n}^{i}(MH) \right] =0; \nonumber\\ & \qquad \forall {{n} \in {\mathcal N}^{i}_{r}}, i\in \mathcal O, r \in \mathcal R.
 \end{align}
\end{lemma}

\begin{proof}
Dividing \eqref{before limsup3} by $\frac{\epsilon}{2}p$, we get

\begin{align}
 & \frac{1}{M} \sum_{k=0}^{MH-1}\mathbb{E} \left[ \sum_{r \in \mathcal {R}} \sum_{i\in \mathcal O}\sum_{{n} \in {\mathcal N}^{i}_{r}} P_{r,n}^{*i}(k) A^{i}_{r,n}(k) \right]  \nonumber \\ & \leq  \frac{B+D}{\frac{\epsilon}{2} p}  + \frac{\eta C}{\epsilon p} -\frac{E\left [ {L}(\Theta^*(MH))\right ]}{M \frac{\epsilon}{2} p}   \nonumber \\
 &  + \frac{V}{\frac{\epsilon}{2} p M}\sum_{k=0}^{MH-1}  E \left [\sum_{r \in \mathcal {R}} \sum_{i\in \mathcal O}\sum_{{n} \in {\mathcal N}^{i}_{r}} Q\left(\tau_{r,n}^{*i}(\bm{S}^*_{t(k)},k)\right)A^{i}_{r,n}(k) \right] \nonumber \\ & - \frac{V}{\frac{\epsilon}{2} p } E \left [\sum_{r \in \mathcal {R}} \sum_{i\in \mathcal O}\sum_{{n} \in {\mathcal N}^{i}_{r}} \sum_{k' \in \mathcal{P}_{t(k)}} Q\left(\tau_{r,n}^{**i}(\bm{S}^{**},\bm{A})\right)A^{i}_{r,n}(k') \right] 
 \end{align}

 Using the fact that $P_{r,n}^{*i}(k)\ge 0$ and $ {L}(\Theta^*(MH)) \geq 0$,

Since all the terms on the right hand side  are bounded, we can conclude that $ limsup_{M \to \infty}$ of the terms on the left hand side are bounded.

\begin{align}
 & \limsup_{M \to \infty} \frac{1}{M} \sum_{k=0}^{MH-1}  E \left[  \sum_{r \in \mathcal {R}} \sum_{i\in \mathcal O}\sum_{{n} \in {\mathcal N}^{i}_{r}}     P_{r,n}^{*i}(k) A_{r,n}^{i}(k)\right ]  \leq G
 \end{align}
where $G$ is a positive constant.
Thus $limsup_{M \to \infty} \frac{1}{M} \sum_{k=0}^{MH-1}  E \left[      P_{r,n}^{*i}(k) A_{r,n}^{i}(k)\right ]$ is also bounded for $\forall {{n} \in {\mathcal N}^{i}_{r}}, i\in \mathcal O, r \in \mathcal R$.
 
Since the above condition holds, we can use Lemma \ref{prk_lemma1} (from \cite{prk}), and Lemma \ref{lemma_new0} follows.

  \end{proof}

\begin{lemma}\label{1 constraint satisfy lemma}
For any strictly static feasible $Q_{min}$, the approximated constraint is satisfied.
\end{lemma}

\begin{proof}
Consider inequality obtained from \eqref{update eqn} and averaging it over $|\mathcal{K}_n|$ periods, we have
\begin{align}
  \frac{ \sum_{k \in \mathcal{K}_n} P_{r,n}^{*i}(k+1)}{|\mathcal{K}_n|} & \geq  \frac{\sum_{k \in \mathcal{K}_n} P_{r,n}^{*i}(k)}{{|\mathcal{K}_n|}} \nonumber \\ & +  \frac{1}{|\mathcal{K}_n|} \sum_{k \in \mathcal{K}_n}\left [ X^{*i}_{r,n}(\bm{S}^*_{t(k)},k) - p \right], \nonumber \\   & \hspace{1.5cm} \forall n \in \mathcal{N}^{i}_{r}, \ i\in \mathcal{O}, \ r \in \mathcal{R} 
\end{align}

Cancelling out common $P_{r,n}^{*i}(k)$ terms and noting that $P_{r,n}^{*i}(0) \ge 0$, we have

\begin{align}
  \frac{ P_{r,n}^{*i}(|\mathcal{K}_n|)}{|\mathcal{K}_n|} \geq   & \frac{1}{|\mathcal{K}_n|} \sum_{k \in \mathcal{K}_n}  \left [X^{*i}_{r,n}(\bm{S}^*_{t(k)},k) - p \right], \nonumber \\ &  \qquad \forall n \in \mathcal{N}^{i}_{r}, \ i\in \mathcal{O}, \ r \in \mathcal{R} \label{before applying lemma3}
\end{align}

Using \eqref{before applying lemma3} and Lemma \ref{lemma_new0}, we thus have
\textcolor{black}{
\begin{align}
 &  \limsup_{|\mathcal{K}_n| \to \infty}\frac{1}{|\mathcal{K}_n|} \sum_{k \in \mathcal{K}_n}  (Q_{min}-Q(\tau^{*i}_{r,n}(\bm{S}^*_{t(k)},k))+\alpha)_+ \leq  0.05\alpha, \nonumber \\ & \qquad \forall n \in \mathcal{N}^{i}_{r}, \ i\in \mathcal{O}, \ r \in \mathcal{R} 
\end{align}}
Thus, ABS satisfies \eqref{modified_convexified_constraint}.

\end{proof}

Note that Theorem \ref{theorem_feasibility_optimal} has now been proven using the feasibility and optimality results in Lemma \ref{1 constraint satisfy lemma} and Lemma \ref{average quality lemma3} respectively.

\end{appendices}}%
  }
  {
   \newcommand{\result}{}%
  }

\begin{document}

\title{Adaptive Bandwidth Sharing for Optimizing QoE of Real-Time Video}

\author{Sushi~Anna~George, Vinay~Joseph
\thanks{S. George and V. Joseph are with the National Institute of Technology, Calicut, India.}
}


\maketitle


\begin{abstract}
The concept of spectrum or bandwidth sharing has gained significant global attention as a means to enhance the efficiency of real-time traffic management in wireless networks. Effective bandwidth sharing enables optimal utilization of available resources, reducing congestion and improving QoE for delay-sensitive applications such as real-time video transmission. In this paper, we propose a novel iterative semi-static bandwidth sharing policy that balances the advantages of both static and dynamic sharing approaches. Our approach minimizes the frequency of coordination between network operators while ensuring efficient resource allocation and meeting the stringent QoE demands of real-time traffic.
The proposed policy iteratively optimizes both the spectrum sharing between operators and the resource allocation for individual clients.
We establish strong theoretical guarantees for the optimality of the proposed policy and prove that it converges to the optimal static sharing policy irrespective of initial conditions or fluctuations in traffic arrival rates. Additionally, we conduct extensive simulations to evaluate the impact of key system parameters—including step size, hyperperiod length, and arrival process dynamics—on the performance of our policy. Our results demonstrate the effectiveness of the proposed approach in achieving near-optimal bandwidth allocation with reduced overhead, making it a practical solution for real-time wireless applications.
\end{abstract}

\begin{IEEEkeywords}
Quality of Experience, QoE, Bandwidth sharing, Real-time traffic, Scheduling, Allocation, Perceived video quality
\end{IEEEkeywords}

\IEEEpeerreviewmaketitle

\section{Introduction}

\IEEEPARstart{T}{he} widespread use of applications such as video and audio conferencing, live event streaming, augmented and virtual reality, and online gaming has driven a significant increase in real-time traffic  \cite{gmi}, \cite{meti_research}. Managing this type of traffic demands special attention, as it is highly sensitive to delays and must be delivered within strict deadlines to maintain sufficient quality.

Bandwidth sharing is a solution to better manage real-time traffic by efficient utilization of the available resources. It enables networks to handle a growing number of users and devices by efficiently managing the available resources. Bandwidth sharing also helps alleviate network congestion, optimize resource allocation, and enhance overall system performance, particularly in environments with limited capacity and fluctuating traffic patterns. This paper proposes an adaptive  bandwidth sharing strategy in wireless networks for real-time video transmission, focusing on reducing the frequency of coordination between operators while maintaining efficient resource utilization and meeting the quality demands of real-time traffic. This approach is able to adapt to the variations in network conditions such as client arrival rates, channel capacities, traffic demands, etc.

\begin{figure}[htbp]
\centerline{\includegraphics[scale=0.25]{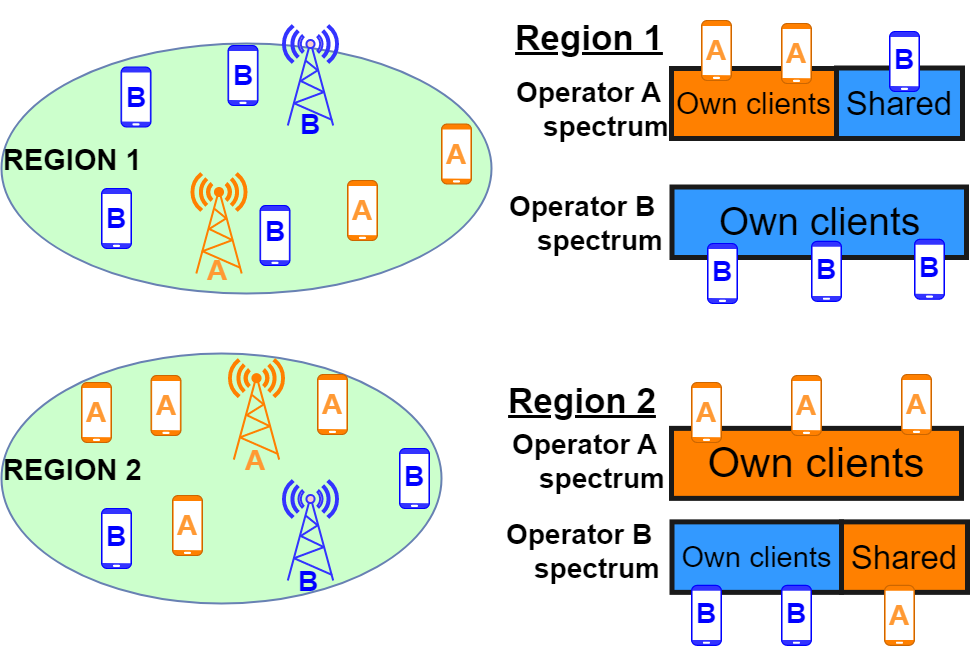}}
\caption{Illustration of bandwidth sharing between operators across regions}
\label{sharingillustration}
\end{figure} 

A form of bandwidth sharing is illustrated in Fig. \ref{sharingillustration}. 
The bandwidth sharing mechanism enables an operator $A$ with lower real-time traffic in region 1 to share some of its timeslots with another operator $B$ in the same region. \\

The idea of spectrum or bandwidth sharing has garnered significant attention globally, with active discussions taking place in various forums, including government agencies (see \cite{policy_impact}, \cite{ofcom}, \cite{nist}, \cite{deptindia}). Bandwidth sharing can be implemented at different time-scales depending on how frequent resources are required to be allocated in different applications. Given the dynamic nature of real-time traffic, resource allocation must adapt to fluctuating traffic peaks and user demands. Hence, dynamic/time-varying bandwidth sharing is essential as it allows for a flexible, demand-responsive approach to resource management, ensuring that network resources are used efficiently and effectively across varying conditions. Dynamic bandwidth allocation determines the demands for each client at every instant and allocates bandwidth accordingly. Dynamic bandwidth sharing requires continuous monitoring of network conditions, user demands, and application requirements. This necessitates real-time data analysis and decision-making processes, which can be computationally intensive and requires frequent inter-operator coordination.

Therefore, in this work, we propose an \textit{adaptive bandwidth sharing} scheme that leverages the advantages of both static and dynamic bandwidth sharing approaches. Our proposed method maintains a \textit{semi-static} structure where the bandwidth shared between operators remains constant throughout a hyperperiod (a group of periods), reducing the need for frequent inter-operator coordination. The bandwidth allocation is then updated adaptively at the start of the next hyperperiod based on observed packet arrivals, allowing the system to respond dynamically to traffic variations while ensuring stability and efficient resource utilization. This approach strikes a balance between minimizing overhead and adapting to network conditions, making it well-suited for real-time traffic scenarios. 

\subsection{Main Contributions} \label{key contributions}

{Major contributions} of this work are summarized below:

\begin{enumerate} 
\item We introduce an adaptive semi-static bandwidth sharing policy for multiple operators, aimed at maximizing the perceived video quality for clients and reducing the frequency of coordination between operators (see Section \ref{sectionsemisharing}). 
\item We demonstrate that the performance of the semi-static bandwidth sharing policy converges to that of the optimal static bandwidth sharing policy through simulations and provide strong theoretical optimality guarantees for our policy (see Theorem \ref{theorem_feasibility_optimal}).
\item We assess the effectiveness of the optimal static bandwidth sharing policy by comparing its performance to that of the optimal dynamic bandwidth sharing policy through simulations (see Section \ref{sectionperformanceevaluation}). The results show that the optimal static bandwidth sharing policy achieves performance nearly equivalent to the optimal dynamic approach. 
\item We investigate the impact of various parameters like stepsize, hyperperiod length, etc on the convergence of the semi-static bandwidth sharing policy through extensive simulations and provide detailed insights (see Section \ref{sectionperformanceevaluation}).
\end{enumerate}

\subsection{Organization of the paper} \label{organisation}
We discuss related work in Section \ref{section_related_work}.
System model is discussed in Section \ref{sectionsystemmodel}. The formulations for optimal static and optimal dynamic bandwidth sharing policies are briefed in Section \ref{Od OS}. Section \ref{sectionsemisharing} presents the proposed semi-static bandwidth sharing policy. Simulation results are in Section \ref{sectionperformanceevaluation}.
We conclude in Section \ref{sectionconclusion}.

\section{Related Work} 
\label{section_related_work}

\subsection{Relevance of bandwidth sharing and its implementation}
Spectrum leasing can be viewed as bandwidth sharing (at a slow time scale), and is used in a wide range of use-cases including those involving mobile virtual
network operators (MVNOs) and private networks (\cite{fcc_leasing}, \cite{gsma_leasing}). Spectrum sharing is an important focus area for Citizens Broadband Radio Service (CBRS) in USA, with spectrum re-allocation possible in roughly 5 minutes \cite{fccCBRS}.

Bandwidth sharing also requires tight \textit{inter-operator coordination}. Techniques in 4G/5G like Multi-Operator Core Network (MOCN), Gateway Core Network (GWCN) and Multi-Operator Radio Access Network (MORAN) enable multiple operators to coordinate and use shared network infrastructure (\cite{3gpp_ts_32_130}, \cite{ericsson_japan}, \cite{ericsson_taiwan}). Progress in solutions leveraging Software Defined Networking (SDN) and Network Function Virtualization (NFV) \cite{huawei_nfv} also make it easier to share network/compute resources, and thus allow flexible and tighter coordination among operators. There are deployment-ready solutions now like 5G Radio Dot System \cite{ericsson_radioDot}, which support  service by multiple operators. Despite the significant interest in spectrum sharing, there has been very little work exploiting spectrum sharing for real-time traffic, and none of them consider perceived video quality or Quality of Experience (QoE).
We address this gap in this paper.

\subsection{Infrequent decision making}
Implementing per period spectrum sharing between operators is indeed challenging due to the dynamic nature and the real-time coordination required. Many papers consider the solution of making infrequent decisions. Even though \cite{kashef2015exploiting} claims that the performance for the case of making decision
every time slot is better than the case of infrequent decision, the performance improvement is very low (about 3.5\%). \cite{gopalan2013value} presents a throughput optimal scheduling algorithm that relies on the base stations sharing their users’ queue lengths every $T$ time slots and for the next $T$ time slots, each base station uses this (delayed) queue length information. The work in \cite{neely} demonstrates that infrequent queue updates, as long as the maximum per-timeslot backlog change in any queue is bounded, do not compromise the stability of the system.

Infrequent decision making is also evident in situations where a decision made is delayed by a few timeslots. In many wireless systems, Channel State Information (CSI) is only available after a sequence of channel estimation, quantization, and feedback processes. \cite{wang2022periodic} formulates a new constrained Online Convex Optimization (OCO) problem with periodic updates with each update period lasting for multiple time slots.

\subsection{On scheduling of real-time traffic in wireless networks}
There has been previous work on providing services for delay-constrained traffic in wireless networks. One of our key references is  \cite{hou2009theory}, which analyzed scheduling real-time traffic in unreliable wireless environments. \cite{hou2009theory} introduced a term \textit{timely throughput} to measure the amount of real-time traffic that is successfully delivered. Further, \cite{hou2013scheduling} and \cite{prk} developed scheduling policies with optimality guarantees for real-time traffic for various scenarios (e.g., rate adaptation, time-varying channels). Scheduling real-time traffic with hard
deadlines in a wireless ad hoc network by ensuring both timely throughput and data freshness guarantees for deadline-constrained traffic is considered in \cite{lu2018age}.
The system models in \cite{hou2009theory}, \cite{hou2013scheduling},\cite{prk} and \cite{lu2018age} rely on \textit{frame-based models} for arrival and scheduling of real-time traffic. In frame-based models (which is used in this paper too), all traffic arrives at the beginning of a frame and has to be served by the end of the frame. 
The approach in \cite{tsanikidis2021power} does not require this frame-based approach and uses an approach relying on randomization of the choice of transmitting links. Papers like \cite{age_optimal_sun}, \cite{timely_qi} have also approached this topic by formulating the problem in terms of \textit{age of information} and developed scheduling algorithms using deep reinforcement learning.
Note that these previous works do not however explore solutions leveraging bandwidth sharing.
\cite{ICC_paper} proposes a solution employing bandwidth sharing for serving real-time traffic, though it uses a simplified system model with a simplified resource allocation model and ignores perceived video quality.

\subsection{Optimizing QoE for wireless network} 
\cite{huang_joint_source}, \cite{fu2010systematic}, and \cite{joseph2012jointly} propose network optimization to maximize metrics reflecting QoE or perceived video quality of users in a wireless setting. 
\cite{huang_joint_source} and \cite{fu2010systematic} use PSNR metric for perceived video quality whereas \cite{joseph2012jointly} considers a more general QoE metric modelled as a function of PSNR and SSIM values over time. However, these works too do not consider bandwidth sharing.
\section{System Model}\label{sectionsystemmodel}

\begin{table}[]
\caption{Notation}
    \centering
    \begin{tabular}{p{0.06\textwidth} p{0.38\textwidth}}
    \hline
\textbf{Notation}  & \textbf{Description} \\
\hline
\hline
 ${\mathcal O}$    &      Set of all operators\\
        \\[-1em]
 $i, j$      &    Indices used for operators\\
        \\[-1em]
${\mathcal R}$ &     Set of all regions
\\
        \\[-1em]
$r$     &       Index used for a region\\
        \\[-1em]
 ${\mathcal N}^{i}_{r}$ & Set of all clients in the $r^{th}$ region of the $i^{th}$ operator\\
  \\[-1em]
 ${\mathcal N}^{i}_{r}(k)$ & Subset of ${\mathcal N}^{i}_{r}$ comprising of clients with packet arrival in period $k$\\
        \\[-1em] 
        $n$ &   Index used for a client\\ 
        \\[-1em]
             $\mathcal{K}$ &  Set of all periods\\
        \\[-1em]
         $k$ &  Index of a period\\         
        \\[-1em]
        $T$ & Number of timeslots within a period\\
         \\[-1em]
             $H$ &  Number of periods within a hyperperiod\\
        \\[-1em]
         $\mathcal{P}_t$ & Set of all periods within the $t^{th}$ hyperperiod\\        
        \\[-1em]
         $ A^{i}_{r,n}(k)$  & Number of packet arrivals of $n^{th}$ client in the $r^{th}$ region of the $i^{th}$ operator in the $k^{th}$ period\\
      \\[-1em]
      $\rho^i_{r,n}$ & Average number of packet arrivals per period for $n^{th}$ client in the $r^{th}$ region of the $i^{th}$ operator\\
      \\[-1em]
       $ A^{i}_{r}(k)$  & Number of packet arrivals across all clients in the $r^{th}$ region of the $i^{th}$ operator in the $k^{th}$ period\\
       \\[-1em]
       $c_{r,n}^{i}(k)$ & Channel capacity of the $n^{th}$ client in $r^{th}$ region of $i^{th}$ operator (number of bits that can be delivered to the client in one timeslot) in the $k^{th}$ period\\
      \\[-1em]
     $\tau_{r,n}^{i}(k)$ & Number of timeslots alloted to  the $n^{th}$ client in $r^{th}$ region of $i^{th}$ operator in the $k^{th}$ period \\
       \\[-1em]
       $Q_{r,n}^{i}(.)$ & Perceived video quality of the $n^{th}$ client in $r^{th}$ region of $i^{th}$ operator \\
       \\[-1em]
       $Q_{min}$ & Minimum acceptable perceived video quality\\
         \\[-1em]
         ${S^{j\rightarrow i}_{r}(k)}$   & Number of timeslots of the $j^{th}$ operator shared with the $i^{th}$ operator in the $r^{th}$ region in the $k^{th}$ period\\
         \\[-1em]
         $\zeta^{(i,j)}$ & Bound on difference in sharing of any two operators\\
         \hline
         \hline
    \end{tabular}
    \label{tab:my_label}
\end{table}

We consider the downlink of a wireless system spanning a set of regions ${\mathcal R}$ served by a set of operators $\mathcal O$. 
Let ${\mathcal N}^{i}_{r}$ denote the set of clients in region $r$ of operator $i$. Time is partitioned into timeslots, with $T$ consecutive timeslots comprising a \textit{period}.

\textbf{Wireless channel:} 
We consider a time varying wireless channel model. We denote channel capacity of the $n^{th}$ client in the $r^{th}$ region of the $i^{th}$ operator in the $k^{th}$ period by $c_{r,n}^{i}(k)$.  We assume that $\left\{ c^{i}_{r,n}(k):k\ge 1\right\}$ are independent and identically distributed, and that they are independent for any two clients. Our model is capable of accommodating heterogeneous channel conditions and client-specific settings.

\textbf{Interference model:} 
We assume that an operator is capable of transmitting simultaneously in multiple regions without an interference between the transmissions.

\textbf{Packet arrivals:}
We represent a client’s video traffic as a stream of packets, where packet arrival occurs only at the beginning of each period. We assume that at most one packet arrives for each client in a period. We define $A^{i}_{r,n}(k) \in \{0,1\}$ as an indicator function to determine if a packet arrives for the $n^{th}$ client in the $r^{th}$ region of the $i^{th}$ operator during the $k^{th}$ period. We assume that $\left\{ A^{i}_{r,n}(k):k\ge 1\right\}$ are independent and identically distributed, and that they are independent for any two clients.
We let $\rho^i_{r,n}$ denote the average number of packet arrivals per period for the $n^{th}$ client in the $r^{th}$ region of the $i^{th}$ operator.

\textbf{Resource allocation}:
Let $\tau^{i}_{r,n}(k)$ denote the number of timeslots (i.e., resources) allotted to the $n^{th}$ client in the $r^{th}$ region of the $i^{th}$ operator in the $k^{th}$ period.
We assume that timeslots are allotted only to a client with a packet arrival in period $k$, i.e., $\tau_{r,n}^{i}(k) =0$ if $ A_{r,n}^{i}(k)=0$.
Other forms of resources like subcarriers in an OFDMA network (e.g., 4G, 5G) can also be considered instead of timeslots.

\textbf{Quality of Experience (QoE): } 
 We model  perceived video quality of the $n^{th}$ client in $r^{th}$ region of $i^{th}$ operator in the $k^{th}$ period as a function $Q^{i}_{r,n}(\tau_{r,n}^{i}(k))$ of $\tau_{r,n}^{i}(k)$. 
 We assume that $Q^{i}_{r,n}(.)$ is a twice-differentiable increasing concave function of the number of timeslots allotted to the client. 
Such concavity is observed in various video quality metrics like PSNR (Peak Signal-to-Noise Ratio) \cite{huang_joint_source}, MOS (Mean Opinion Score)  \cite{cermak2011relationship}, and (negative of) distortion functions \cite{cover1999elements}. Client-specific quality functions can also be incorporated to better reflect real-world scenarios. 
To simplify the notation, we will use $Q^{i}_{r,n}(k)$ instead of $Q^{i}_{r,n}(\tau_{r,n}^{i}(k))$, for discussions where it is not critical to highlight its  dependence on $\tau_{r,n}^{i}$.

MOS metric provides a standardized way to evaluate user satisfaction and perceived quality across different scenarios, making it a widely used benchmark for comparing performance in applications like video streaming, voice, and other multimedia services \cite{cermak2011relationship}.

The perceived video quality across periods dictates the \textit{Quality of Experience (QoE)} of a client. We model the QoE of a client as the average of video quality perceived by the client across all periods. 

\textbf{Quality percentile guarantee:} A defining feature of real-time video is that each packet must be delivered within a specific deadline; otherwise, the packet loses its value or becomes significantly less useful. 
For real-time traffic, timely delivery (i.e., delivery within deadline) of packets plays a critical role in determining user experience.  The quality of experience (QoE) for users is highly dependent on timely delivery, as excessive delays lead to video freezing, stalling, or degradation in video quality. 

To ensure an acceptable level of QoE, we impose a constraint that guarantees a minimum proportion of packets meet a predefined quality threshold. Specifically, we require that at least 95\% of the packets for each client $n$ meet or exceed a minimum acceptable perceived video quality $Q_{min}$. This can be expressed as:

\begin{align}
   & \mathbb{E} \left[\mathds{1} \left\{ Q^{i}_{r,n}(k) \geq Q_{min}\right\}\right] \geq 0.95; \nonumber \\ & \hspace{2.5cm} \forall{n} \in {\mathcal N}^{i}_{r},  \forall {i} \in {\mathcal O},  \forall{r} \in {\mathcal R}, \forall{k} \in \mathcal{K}_n \label{indicator constraint}
\end{align}
where $\mathds{1} \left\{ Q^{i}_{r,n}(k) \geq Q_{min}\right\}$  is an indicator function that takes the value 1 if the quality of packet in the $k^{th}$ period meets or exceeds $Q_{min}$, and 0 otherwise. $\mathcal{K}_{n}= \left\{k \in \mathcal{K} | n \in \mathcal{N}^{i}_{r}(k) \right\}$ denotes the set of all periods where client $n$ has packet arrivals and $K_n=|\mathcal{K}_{n}|$.
We consider the indicator constraint above without loss of generality, as this approximation aligns with the recommended 5\% permissible packet loss from \cite{webex}. \eqref{indicator constraint} can be considered similar to enforcing that the $5^{th}$ percentile of quality of all packets of a client over all periods to be at least $Q_{min}$. 




\subsection{Bandwidth Sharing Model} 
The bandwidth sharing framework provides an insight into how resources in terms of timeslots within a period, can be shared among operators in a region. Let 
${S^{j\rightarrow i}_{r}(k)}$ denote number of timeslots operator $j$ shares with operator $i$ in region $r$ during period $k$. Note that 
${S^{i\rightarrow i}_{r}(k)}$ denotes the number of timeslots operator $i$ uses for its own clients in region $r$ during period $k$.
Next, we discuss the constraints associated with bandwidth sharing variables.

Clearly, sharing-related variables need to be non-negative and thus we have
\begin{align}
 \qquad \quad S^{j\rightarrow i}_{r}(k) \geq 0,  \qquad \forall {i,j} \in {\mathcal O}, \forall{r} \in {\mathcal R}, \forall{k} \in {\mathcal K}.\label{nonneg} 
\end{align}
Also, the below constraint captures that maximum number of timeslots available to an operator for its own use and for sharing with other operators is $T$:
\begin{align}
& \qquad \sum_{i\in {\mathcal O}}S^{j\rightarrow i}_{r}(k) \leq T, \qquad  \forall {j} \in {\mathcal O}, \forall{r} \in {\mathcal R}, \forall{k} \in {\mathcal K}. \label{maxshare} 
\end{align}

Without sharing, the number of timeslots that can be allotted by operator $i$ to its clients in region $r$ is bounded by ${S^{i\rightarrow i}_{r}(k)}\leq T$.
With sharing, the number of timeslots that can be allotted by operator $i$ to its clients in region $r$  is bounded by the sum of ${S^{i\rightarrow i}_{r}(k)}$ (timeslots owned by the operator) and
$\sum_{j\in {\mathcal O}\setminus \{i\}}S^{j\rightarrow i}_{r}(k)$ (timeslots shared by other operators), i.e., we have:
 \begin{align}
 \sum_{n\in {\mathcal N}^{i}_{r}}\tau_{r,n}^{i}(k) & \leq \sum_{j\in {\mathcal O}}S^{j\rightarrow i}_{r}(k), \ \forall {i} \in {\mathcal O}, \forall{r} \in {\mathcal R}, \forall{k} \in {\mathcal K}.  \label{maxschedule} 
 \end{align}

In order to incentivize sharing between operators, here we ensure that any two operators share the same net amount of bandwidth across all regions and this is captured in the constraint below:
\begin{align}
 \text{Prob} \left\{ \hspace{-0.1cm} \frac{1}{K} \left | \sum_{k=1}^{K} \sum_{r \in \mathcal {R}} \hspace{-0.05cm} S^{j\rightarrow i}_{r}(k)  - \hspace{-0.1cm}\sum_{k=1}^{K} \sum_{r \in \mathcal {R}} S^{i\rightarrow j}_{r}(k)\right |\hspace{-0.05cm} \leq \zeta^{(i,j)} \hspace{-0.1cm} + \hspace{-0.05cm}\xi_2 \right\} \nonumber \\  \rightarrow \text{1},   \text{~as~} K\rightarrow\infty, \label{sharing constraint} 
\end{align}
where $\xi_2$ is a small positive constant. 
The parameter $\zeta^{(i,j)}$  governs the extent of sharing between operators $i$ and $j$. When 
$\zeta^{(i,j)}$  is small, \eqref{sharing constraint} effectively ensures that operators contribute roughly the same amount of resources as they receive from each other across all regions. This constraint encourages balanced sharing (operators only receive as much as they provide) while preventing excessive sharing (operators only provide as much as they receive). This helps create a more balanced and fair system that encourages participation. 

 \section{Optimizing Quality of Experience (QoE)}
 \label{Od OS}

We frame the problem of optimizing system-wide QoE as an optimization problem (OPT) below:
 \begin{align}
& \underset{\substack{ \tau_{r,n}^{i}(k), \ S^{j\rightarrow i}_{r}(k) , \\ \forall{n} \in {\mathcal N}^{i}_{r}, \ \forall {i} \in {\mathcal O}, \\ \forall{r} \in {\mathcal R}}}{\text{max}}
 \frac{1}{K} \sum_{k=1}^{K} \sum_{r \in \mathcal {R}} \sum_{i\in \mathcal O} \sum_{{n} \in {\mathcal N}^{i}_{r}} Q^{i}_{r,n}(\tau_{r,n}^{i}(k))A_{r,n}^{i}(k) 
\label{optss'_objective}
\\
\
& \qquad \text{s.t.} \hspace{2cm}  \eqref{indicator constraint}- \eqref{sharing constraint}, \nonumber
\end{align}

Two main decision variables involved in the above optimization problem are:
\begin{itemize}
    \item how many timeslots are allotted to each client, i.e., deciding
    $\bm{\tau}(k) = \left( \tau_{r,n}^{i}(k): \ \forall{n} \in {\mathcal N}^{i}_{r}, \ \forall{r} \in {\mathcal R}, \  \forall {i} \in {\mathcal O} \right) $ for each $k$, and
    \item how much to share, i.e., deciding $\mathbf{S}(k) = \left(S^{j\rightarrow i}_{r}(k): \ \forall{r} \in {\mathcal R}, \  \forall {i,j} \in {\mathcal O}\right)$ for each $k$. 
\end{itemize}

In the sequel, we discuss the three approaches listed below for solving OPT:
\begin{enumerate}
    \item Optimal dynamic sharing policy: requiring frequent inter-operator coordination; 
    \item Optimal static sharing policy (OPT-SS): requiring only limited inter-operator coordination; 
    \item Approximation of optimal static sharing policy (OPT-SS*): a convex relaxation of optimal static sharing policy; 
    \item An iterative semi-static sharing policy: a practical way of realizing the approximation of optimal static sharing policy. 
\end{enumerate}


\subsection{Optimal Dynamic-sharing Policy}  

Optimal dynamic-sharing policy can be obtained if we solve OPT. Solving it however is not straightforward, as it includes constraints \eqref{indicator constraint} and \eqref{sharing constraint}.
We solve OPT in Section \ref{sectionperformanceevaluation}.
Applying the approach in \cite{arxivpaper2024} to real-world networks essentially involves  updating sharing of bandwidth/spectrum among operators in real-time based on real-time information (of arrivals). 
This requires coordination and exchange of information between operators every period. This level of frequent coordination can be challenging in practice, and the next policy addresses this limitation.




\subsection{Optimal Static-sharing Policy} 
Here, we consider a class of policies in which bandwidth sharing among operators remains fixed or \textit{static}.
We let ${S^{j\rightarrow i}_{r}}$ denote the 
 number of timeslots operator $j$ shares with operator $i$ in region $r$ \textit{in each period}. That is, ${S^{j\rightarrow i}_{r}}$ is the static variant of ${S^{j\rightarrow i}_{r}}(k)$ defined earlier.
Let $\bm{S}$ be a vector with static sharing variables ${S^{j\rightarrow i}_{r}}$ as its components.

The \textit{optimal} static sharing policy can be obtained by solving OPT after restricting the sharing variables to be constant across periods, and we refer to the resulting optimization problem as OPT-SS. 
That is, we obtain OPT-SS by replacing ${S^{j\rightarrow i}_{r}}(k)$ in OPT with ${S^{j\rightarrow i}_{r}}$, and replacing \eqref{sharing constraint} with:
\begin{align}
  \left|\sum_{r \in \mathcal {R}} S^{j\rightarrow i}_{r}  -  \sum_{r \in \mathcal {R}} S^{i\rightarrow j}_{r}\right | \leq \zeta^{(i,j)}  ,  \quad \forall {i,j} \in {\mathcal O}.  \label{sharing constraint_policy}
\end{align}

 Since the bandwidth sharing does not adjust to fluctuating packet arrivals, the optimal static sharing requires much lesser inter-operator coordination. However, the optimal static sharing policy is difficult to implement in practice because it requires complete information about all future arrivals across all operators and regions in advance.  In real-world scenarios, the arrival of users or traffic demands is inherently stochastic and cannot be perfectly predicted. The reliance on complete and accurate information about future arrivals makes the static sharing policy impractical in dynamic, real-world settings where demand patterns are uncertain and continuously changing.

 Further, its inability to adapt to arrivals can limit its performance and this is studied further in Section \ref{subsectionsimulationsettings}.

\subsection{Approximation of Optimal Static Sharing Policy}
Computing optimal solution to OPT-SS is not straightforward due to the presence of the non-convex indicator constraint \eqref{indicator constraint}. 
Hence, we present the following lemma which we use next to convexify  \eqref{indicator constraint}, and its proof is given in Appendix \ref{proof_approx}.

\begin{lemma}\label{approx lemma}
For $\alpha > 0$, enforcing the following constraint
\begin{align}
& \mathbb{E}\left[\left(Q_{min}-Q^{i}_{r,n}(k) + \alpha\right)_+ \right] \leq 0.05 \alpha; \nonumber \\ & \hspace{3cm} \forall n \in \mathcal{N}^{i}_{r}, \ i\in \mathcal{O}, \ r \in \mathcal{R}, \ k \in \mathcal{K}_n  \label{convexified_constraint}
\end{align}
ensures that constraint \eqref{indicator constraint} is satisfied.
\end{lemma} 

Larger $\alpha$ loosens the constraint, allowing more flexibility, while smaller $\alpha$ enforces stricter adherence of $\{Q^{i}_{r,n}(k): k \in{\mathcal K}_n\}$ to $Q_{min}$. The optimal value of $\alpha$ can be determined efficiently using convex optimization methods.

Approximating the expectation in \eqref{convexified_constraint} by corresponding time average, we obtain the following condition, which ensures that \eqref{indicator constraint} is satisfied :
\begin{align}
    &  \frac{1}{K_n} \sum_{k \in \mathcal{K}_n} \left(Q_{min}-Q^{i}_{r,n}(k) +\alpha\right)_+ \leq  0.05 \alpha; \nonumber \\ & \hspace{3cm} \forall n \in \mathcal{N}^{i}_{r}, \ i\in \mathcal{O}, \ r \in \mathcal{R} \label{approx of expectation}
\end{align}

By replacing \eqref{indicator constraint} in OPT-SS with \eqref{approx of expectation},
we obtain a new optimization problem OPT-SS* given below.\\

\hrule
\hrule\vspace{.1cm}
{\begin{center}
    OPT-SS*$_K(\bm{A})$
\end{center} }
\vspace{.1cm}
\hrule
\begin{align}
& \underset{\substack{ \tau_{r,n}^{i}(k), \ S^{j\rightarrow i}_{r}  , \\ \forall{n} \in {\mathcal N}^{i}_{r}, \ \forall {i} \in {\mathcal O}, \\ \forall{r} \in {\mathcal R}}}{\text{max}}
 \frac{1}{K} \sum_{k=1}^{K} \sum_{r \in \mathcal {R}} \sum_{i\in \mathcal O} \sum_{{n} \in {\mathcal N}^{i}_{r}} Q^{i}_{r,n}(\tau_{r,n}^{i}(k))A_{r,n}^{i}(k) \label{problem}
\\
\
& \text{s.t.} \quad
\eqref{sharing constraint_policy}, \ 
\eqref{approx of expectation},
\nonumber
\\
&\qquad \quad S^{j\rightarrow i}_{r} \geq 0,  \qquad \forall {i,j} \in {\mathcal O}, \forall{r} \in {\mathcal R}.\label{nonneg_optss'} 
\\ 
& \qquad \sum_{i\in {\mathcal O}}S^{j\rightarrow i}_{r} \leq T, \qquad  \forall {j} \in {\mathcal O}, \forall{r} \in {\mathcal R}. \label{maxshare_optss'}  
\\ & \sum_{n\in {\mathcal N}^{i}_{r}}\tau_{r,n}^{i}(k)  \leq \sum_{j\in {\mathcal O}}S^{j\rightarrow i}_{r}, \ \forall {i} \in {\mathcal O}, \forall{r} \in {\mathcal R}.  \label{maxschedule_optss'} 
\end{align}
\hrule
Let $(\bm{\tau^*},\bm{S}^*)$  denote the optimal solution to OPT-SS*$_K(\bm{A})$, where $\bm{\tau^*}$ and $\bm{A} $ denote the vectors
comprising of elements $\tau_{r,n}^{*i}(k)$ and $A_{r,n}^{i}(k)$ respectively.

We also consider related optimization problem denoted as OPT-SS*$(\bm{A},\bm{S})$, which is a  restriction of OPT-SS*$(\bm{A})$ obtained by fixing the sharing variables $\bm{S}$ (and only optimizing variables
$\tau_{r,n}^{i}(k)$). 
We will denote its optimal solution as  
$\bm{\tau^*(S)}$
comprising of elements $\tau_{r,n}^{*i}(\bm{S},k)$.

The next result states that OPT-SS* is a convex optimization problem.
 \begin{lemma}
 \label{theorem_concavity}
    OPT-SS*$_K(\bm{A})$ is a convex optimization problem.  
  \end{lemma}
  The  above result is significant, as the convexity now allows for efficient techniques to solve OPT-SS*.  We skip a detailed proof for brevity, and note that proof relies on concavity of objective function, linearity of constraints except \eqref{sharing constraint_policy} and
\eqref{approx of expectation}. Further, \eqref{sharing constraint_policy} can be expressed as two linear constraints and \eqref{approx of expectation} is convex function since maximum of convex functions is convex as well.

\section{Our Proposed Policy: ABS} \label{sectionsemisharing}

The optimal static sharing policy requires less inter-operator coordination compared to the optimal dynamic sharing policy, but its practical implementation is infeasible because it relies on having complete knowledge of all future arrivals in advance—a requirement that is unattainable in real-world scenarios.
 Dynamic sharing, on the other hand, continuously adjusts to network and user conditions but requires real-time monitoring and complex computations. We now present an iterative semi-static sharing policy - \textit{Adaptive Bandwidth-sharing and Scheduling (\textbf{ABS})} which does not require frequent inter-operator coordination inherent in dynamic sharing.

 \textbf{ABS's scheduling}: ABS policy makes bandwidth sharing decisions once every  hyperperiod, which is a collection of consecutive $H$ periods. 
 Let $\mathcal{P}_t$ denote the set of periods associated with the $t^{th}$ hyperperiod.

ABS meets the long-term minimum-quality constraint \eqref{approx of expectation} using a virtual queue $ P_{r,n}^{i}(k)$ associated with \eqref{approx of expectation}. This queue serves to monitor the deficit a client experiences in meeting the requirement specified by \eqref{approx of expectation}.
Let $\bm{P}(k)$ denote an array with entries $P_{r,n}^{i}(k)$ for each $n\in {\mathcal N}^{i}_{r}, \ \forall{r} \in {\mathcal R}, \  \forall {i} \in {\mathcal O}$. 

In each period $k$, ABS carries out scheduling (i.e., deciding $\bm{\tau}$) by optimizing
\begin{align}
  & F_k(\bm{\tau},\bm{P}(k))  =  - V \sum_{r \in \mathcal {R}} \sum_{i\in \mathcal O}\sum_{{n} \in {\mathcal N}^{i}_{r}}  Q \left(\tau^{i}_{r,n}(k) \right)A^{i}_{r,n}(k)  \nonumber \\
  & +\sum_{r \in \mathcal {R}} \sum_{i\in \mathcal O} \sum_{{n} \in {\mathcal N}^{i}_{r}} P_{r,n}^{i}(k)\left(Q_{min}-Q(\tau^{i}_{r,n}(k))+\alpha \right)_+ 
   \label{eqn:fdefn}
\end{align}
where $\bm{\tau}$ is an array with entries $\tau_{r,n}^{i}(k)$ for each $n\in {\mathcal N}^{i}_{r}, \ \forall{r} \in {\mathcal R}, \  \forall {i} \in {\mathcal O}$ 
and  $V$ is a positive constant impacting performance. Let the sharing variables of our policy in hyperperiod $\mathcal{P}_t$ be denoted by $S^{j\rightarrow i}_{t,r}$ and $\mathbf{S}_{t}$ be the vector composed of $S^{j\rightarrow i}_{t,r}$ $\forall {i,j} \in {\mathcal O}, \forall{r} \in {\mathcal R}$.
Specifically, ABS determines scheduling $\bm{\tau}^*(\bm{S}_t^*,k)$ for period $k$ of hyperperiod $\mathcal{P}_t$
by solving the following optimization problem RA$(\bm{S}_t,\bm{A}(k))$.\\
\hrule
RA$(\bm{S}_t,\bm{A}(k))$
\hrule
\begin{align}
& \underset{\substack{ \ \bm{\tau} }}{\text{min}} 
& &    F_k(\bm{\tau},\bm{P}(k))
\label{obj_static}\\
\
& \qquad \text{s.t.} & &  
 \sum_{n\in {\mathcal N}^{i}_{r}}\tau_{r,n}^{i}  & \leq \sum_{j\in {\mathcal O}}S^{j\rightarrow i}_{t,r} , \ \forall {i} \in {\mathcal O}, \forall{r} \in {\mathcal R}.
 \label{maxschedule_staticsharing} 
\end{align}
\hrule
Let $\lambda_{r}^{i}(\bm{S}_t,\bm{A}(k))$ denote the optimal Lagrangian multiplier corresponding to constraint \eqref{maxschedule_staticsharing} for operator $i$ and region $r$ in RA$(\bm{S}_t,\bm{A}(k))$.

\textbf{ABS's bandwidth-sharing}: ABS adapts bandwidth sharing every hyperperiod using the following update rule:
\begin{align}
   \mathbf{S}_{t+1} = \Pi_{\Omega}(\mathbf{S}_t - \eta \textbf{g}(\mathbf{S}_t, \mathbf{A}_{t})) \label{update S}
\end{align}
where $\mathbf{S}_{t}$ is a vector composed of $S^{j\rightarrow i}_{t,r}$ $\forall {i,j} \in {\mathcal O}, \forall{r} \in {\mathcal R}$, in $\mathcal{P}_t$, $\mathbf{A}_{t}$ denote the packet arrivals in $\mathcal{P}_{t}$, and $\textbf{g}(\mathbf{S}_t, \mathbf{A}_{t}) $ is a vector with components
\begin{align}  
\label{g_definition}
g^{j\rightarrow i}_{r}(\mathbf{S}_t, \mathbf{A}_{t})=-\sum_{k \in \mathcal{P}_t}{\lambda}_{r}^{i} 
  (\bm{S}_t,\bm{A}(k)), \ \forall {i, \ j} \in {\mathcal O}, \forall{r} \in {\mathcal R}.
  \end{align}
  \eqref{update S} along with the above update equation ensures that we tend to increase $S^{j\rightarrow i}_{t,r}$ if associated Lagrange multiplier is high. Associated intuition is that the Lagrange multiplier signals the "value" associated with the sharing variable, and we expand on this further in Lemma \ref{gradient in terms of lagrange multipliers}.
  
  $\Pi_{\Omega}$ denotes the projection onto $\Omega$ (as defined in \cite{boyd2004convex}) where $\Omega = \left\{ \bm{S}_t \in \mathbb{R}^n \mid \bm{S}_t \mbox{ satisfies } \eqref{nonneg}, \eqref{maxshare}, \eqref{sharing constraint_policy}\right\}$.
  The update rule for bandwidth sharing \eqref{update S} is essentially stochastic gradient descent (see Lemma \ref{gradient in terms of lagrange multipliers}). Since bandwidth sharing is updated only once in a hyperperiod, the length of the hyperperiod determines frequency of inter-operator coordination required by ABS.

Our iterative semi-static adaptive bandwidth sharing policy is given in Algorithm \ref{alg}.

\RestyleAlgo{ruled}
\begin{algorithm} 
\SetKwInOut{Input}{Input}
\SetKwInOut{Output}{Output}
\caption{Adaptive Bandwidth-sharing and Scheduling (\textbf{ABS})}\label{alg:iter}
\Input{Packet arrivals $\mathbf{A}_{0}$, stepsize $\eta > 0$}
\Output{Optimal sharing vector for all operators and resource allocation for each client}
STEP 0: Initialize the sharing vector  $\mathbf{S}_1$ = $\mathbf{S}_{NO}$ (sharing vector corresponding to no-sharing policy)\;

 \For{Each hyperperiod $\mathcal{P}_t$ where $t = 1, 2, \dots$ }{  
 STEP 1: 
 \For{each period $k$ in hyperperiod $\mathcal{P}_t$,}
{
STEP 1.1: Use resource allocation in period $k$ as 
${\tau}_{r,n}^{i}(\bm{S}_t,\bm{A}(k))$
obtained by solving 
 RA$(\bm{S}_t,\bm{A}(k))$. This can be done on a per-region and per-operator basis.\;
 STEP 1.2: Update $P_{r,n}^{i}(k+1)$ using
 \begin{align}
 & P_{r,n}^{i}(k+1)  = \nonumber \\
  & \begin{cases}
     \max( P_{r,n}^{i}(k)+ \left(Q_{min}-Q(\tau^{i}_{r,n}(\bm{S}_t,k))+\alpha \right)_+ \\ \quad - 0.05\alpha,0 );    \qquad \forall n \in \mathcal{N}^{i}_{r}(k), \ i\in \mathcal{O}, \ r \in \mathcal{R}  \\
    P_{r,n}^{i}(k); \qquad \forall n \in \mathcal{N}^{i}_{r} \setminus \mathcal{N}^{i}_{r}(k), \ i\in \mathcal{O}, \ r \in \mathcal{R} \label{update eqn}
  \end{cases} 
  \end{align} 
 }
 STEP 2: 
   $\mathbf{S}_{t+1} = \Pi_{\Omega}(\mathbf{S}_t - \eta \mathbf{g}(\mathbf{S}_t, \mathbf{A}_{t})) $;
  \\STEP 3: $t \gets t+1$\;
 } \label{alg}
\end{algorithm}

\subsection{Optimality of ABS policy}

We discuss optimality of ABS policy in this section. A policy makes decisions about scheduling and bandwidth sharing subject to the constraints  \eqref{nonneg}, \eqref{maxshare}, \eqref{maxschedule}, and \eqref{sharing constraint_policy}. 
A policy that uses a fixed bandwidth sharing is referred to as a policy with static-sharing.

A policy with static-sharing that takes actions in a randomized manner based solely on the packet arrivals and channel conditions in the current period (i.e., current "state" or the state in the current period) is referred to as an  \textit{stationary randomized policy with static sharing}. 

   Minimum acceptable perceived video quality $Q_{min}$ is said to be \textit{static feasible}, if there exists a stationary randomized policy with static-sharing satisfying \eqref{convexified_constraint}.  
   
\textcolor{black}{
We define the stricter version of \eqref{convexified_constraint} as :
\begin{align}
& \mathbb{E}\left[\left(Q_{min}-Q^{i}_{r,n}(k) + \alpha\right)_+ \right] \leq \kappa 0.05 \alpha; \nonumber \\ & \hspace{3cm} \forall n \in \mathcal{N}^{i}_{r}, \ i\in \mathcal{O}, \ r \in \mathcal{R}, \ k \in \mathcal{K}_n  \label{modified_convexified_constraint}
\end{align} 
where $0<\kappa<1$.}
Minimum acceptable perceived video quality $Q_{min}$ is said to be \textit{strictly static feasible}, if there exists a constant $\kappa$ with $0<\kappa<1$, and a stationary randomized policy
with static-sharing that satisfies \eqref{modified_convexified_constraint}.

A stationary randomized policy with static sharing that maximizes \eqref{problem} asymptotically (as $K\rightarrow \infty$) is referred to as an  \textit{optimal stationary randomized policy with static sharing}. 

The following theorem provides a strong  theoretical performance guarantee for ABS policy. Its proof is in Appendix \ref{feasibilibility_optimal_proof}. 

\begin{theorem}
 \label{theorem_feasibility_optimal} 
  For any strictly static feasible minimum acceptable perceived video quality $Q_{min}$, ABS policy 
  \textcolor{black}{satisfies \eqref{nonneg}, \eqref{maxshare}, \eqref{maxschedule}, \eqref{sharing constraint_policy}, \eqref{convexified_constraint}} and achieves asymptotic value for the objective function  \eqref{problem} that is within $\mathcal{O}\left (1/V \right )$  that of optimal stationary randomized policy with static sharing satisfying \eqref{modified_convexified_constraint}.
  \end{theorem}

  Thus, our policy achieves feasibility whenever possible, and attains total QoE (summed over all clients) close to that of the optimal stationary randomized policy with static sharing. 
Theorem \ref{theorem_feasibility_optimal} also shows that by setting sufficiently large 
$V$, QoE of our policy can be made arbitrarily close to that of the optimal stationary randomized policy with static sharing. Large 
$V$ can lead to lesser emphasis on debts and thus longer virtual queues.
  

 \section{Performance evaluation} \label{sectionperformanceevaluation}
In this section, we study the performance of our iterative semi-static sharing policy via simulations and assess the impact of its parameters.

\subsection{Simulation Settings}\label{subsectionsimulationsettings}
We consider two operators serving two regions (i.e., ${\cal R}=\{1,2\}$) with 30 clients each. To ensure roughly equal sharing between operators, we set $\zeta^{(i,j)}$ as a small value of 0.001. The packet arrival process for each client is modeled using a Bernoulli distribution with parameter $\gamma^i_{r}$, which denotes the average number of packet arrivals per period for each client in region $r\in{\cal  R}$ of operator $i\in {\cal O}$. For all our simulation experiments involving two regions, we set $\gamma^1_{1}$ = $\gamma^2_{2}$ and $\gamma^1_{2}$ = $\gamma^2_{1}$. This ensures that an operator will have a relatively low arrival rate in one region and a high arrival rate in the other. For most plots (unless stated otherwise), the arrival rates for the two operators in a region are set to 0.1 and 0.9, respectively. We set $c_{r,n}^{i}=10 \times 10^6$ bits per timeslot. We set $T=20$ after considering typical values of acceptable latency and typical interarrival time between video frames.

\subsubsection{Tau-Quality relation}
The most commonly used video codec is H.264 \cite{adobe}, \cite{wiki}, \cite{wowza}, \cite{microsoft}, \cite{zoom}, \cite{webex_VC}. Considering the bandwidth to be 10 MHz, the spectral efficiency to be 5 bps/Hz (based on typical values \cite{5GNR}, \cite{coleago}) and based on  Fig. 4 of \cite{cermak2011relationship}, we obtained the following model for perceived video quality (unitless) for  H.264 video codec.

\begin{align}
   Q_{r,n}^{i}(\tau_{r,n}^{i})  =\frac{1}{0.8} \ln \left(\frac{\frac{\tau_{r,n}^{i}c_{r,n}^{i}}{T}+0.1}{0.4}\right) \label{tau_Q}
\end{align}

360p is the lowest standard definition widely used for live streaming, offering a clear viewing experience without noticeable blur on smaller screens such as smartphones, tablets, and older computers or TVs \cite{boris},\cite{resi}.  A bitrate of around 0.4 Mbps is necessary for video with a minimum resolution of 360p \cite{minBRgoogle}, \cite{dacast}. Therefore, we chose 0.3 as the minimum acceptable perceived quality, or $Q_{min}$.

 We maintain $Q_{min}=0.3$ throughout our simulation section for all polices except the no sharing policy. We set $Q_{min}=0$ in the no-sharing scenario for consistency. Even when comparing our policy to a no-sharing policy with $Q_{min}=0$, we observe a significant improvement. Setting $Q_{min}=0.3$ in the no-sharing scenario would prevent the existence of a feasible solution in some cases, as most clients are unable to meet this threshold under a no-sharing policy. In contrast, our semi-static sharing policy successfully ensures that even these clients meet the $Q_{min}$ requirement.


\subsubsection{Parameters for Optimal static sharing}
The constant $\alpha$ is set to 0.008 based on multiple trial-and-error experiments to balance feasibility and quality constraint satisfaction.

\subsubsection{Parameters for Semi-static sharing}
We set the number of periods in one hyperperiod as 20. The stepsize $\eta$ and constant $\alpha$ are set as 0.01 (for all plots except Fig. \ref{vary step size}) and 0.008 respectively.

\begin{table}[]
\caption{Key simulation parameters}
\begin{tabular}{|c|c|}
\hline
\textbf{Simulation Parameter}                      & \textbf{Assigned Value}                                                                                        \\ \hline
Number of operators $({|\mathcal O|})$                                & 2                                                                                                          \\ \hline
Number of regions  $({|\mathcal R|})$                                & 2                                                                                                          \\ \hline
Number of clients per operator per region  $({|\mathcal N}^{i}_{r}|)$        & 30                                                                                                         \\ \hline
Number of timeslots within a period  $(T )$            & 20                                                                                                         \\ \hline
Channel capacity $(c_{r,n}^{i} )$                                  & \begin{tabular}[c]{@{}c@{}}10 $\times 10^6$\\ bits per timeslot\end{tabular} \\ \hline
Minimum acceptable perceived video quality $(Q_{min})$         & 0.3                                                                                                        \\ \hline
Quality scaling factor $(\gamma_Q )$                            & 0.8                                                                                                        \\ \hline
Bound on difference in sharing $(\zeta^{(i,j)} )$ & 0.001                                                
                \\ \hline
 Constant $\alpha $ & 0.008   

                \\ \hline
Stepsize $(\eta) $ & 0.01   

\\ \hline
\end{tabular}
    \label{simulation_table}
\end{table}
\subsection{Simulation Results}
\begin{figure}[htbp]
\centering
\begin{subfigure}[b]{0.55\textwidth}
   \includegraphics[width=0.88\linewidth]{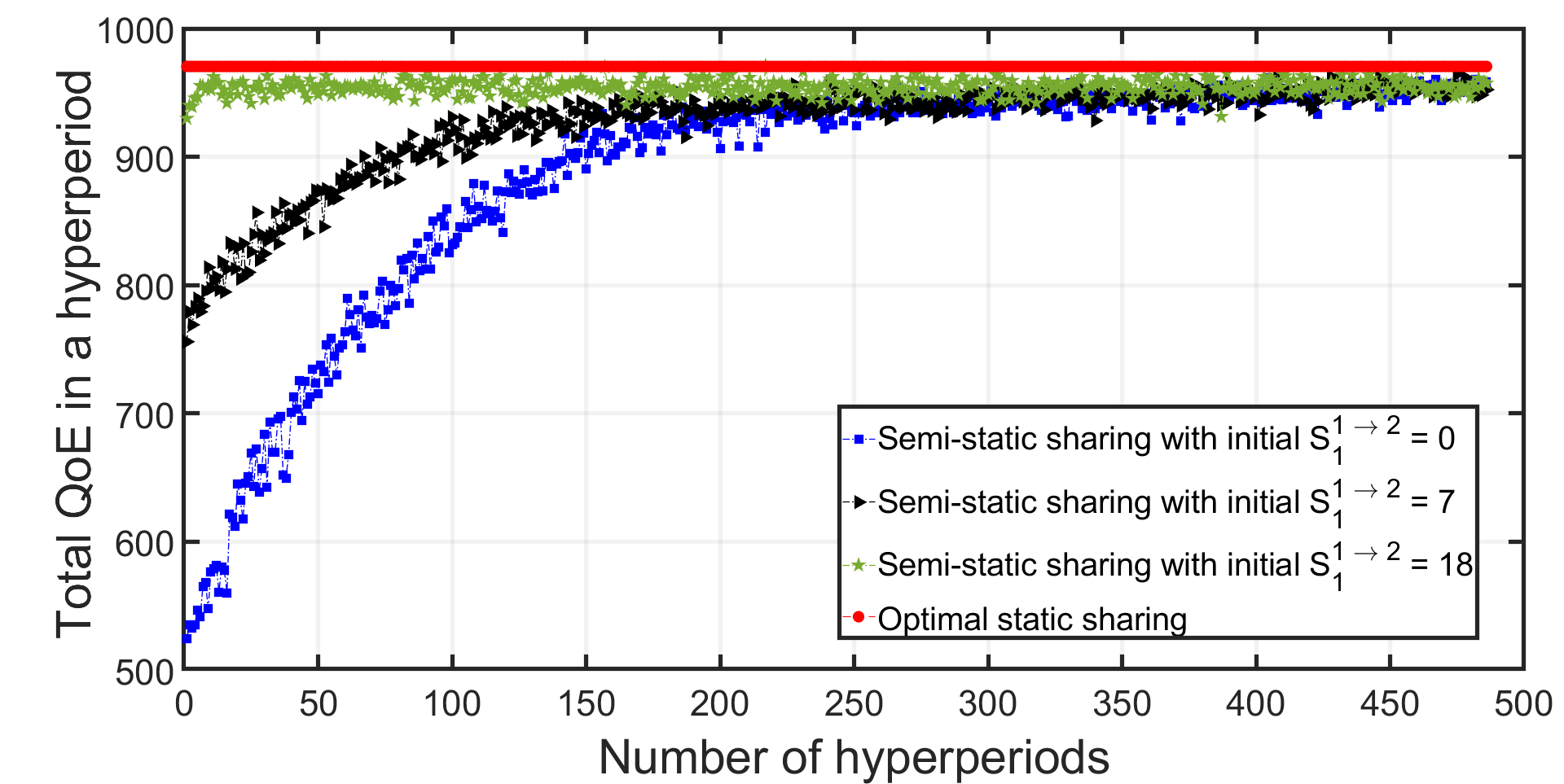}
   \caption{}
   \label{convergence_obj_modified} 
\end{subfigure}
\begin{subfigure}[b]{0.55\textwidth}
   \includegraphics[width=0.88\linewidth]{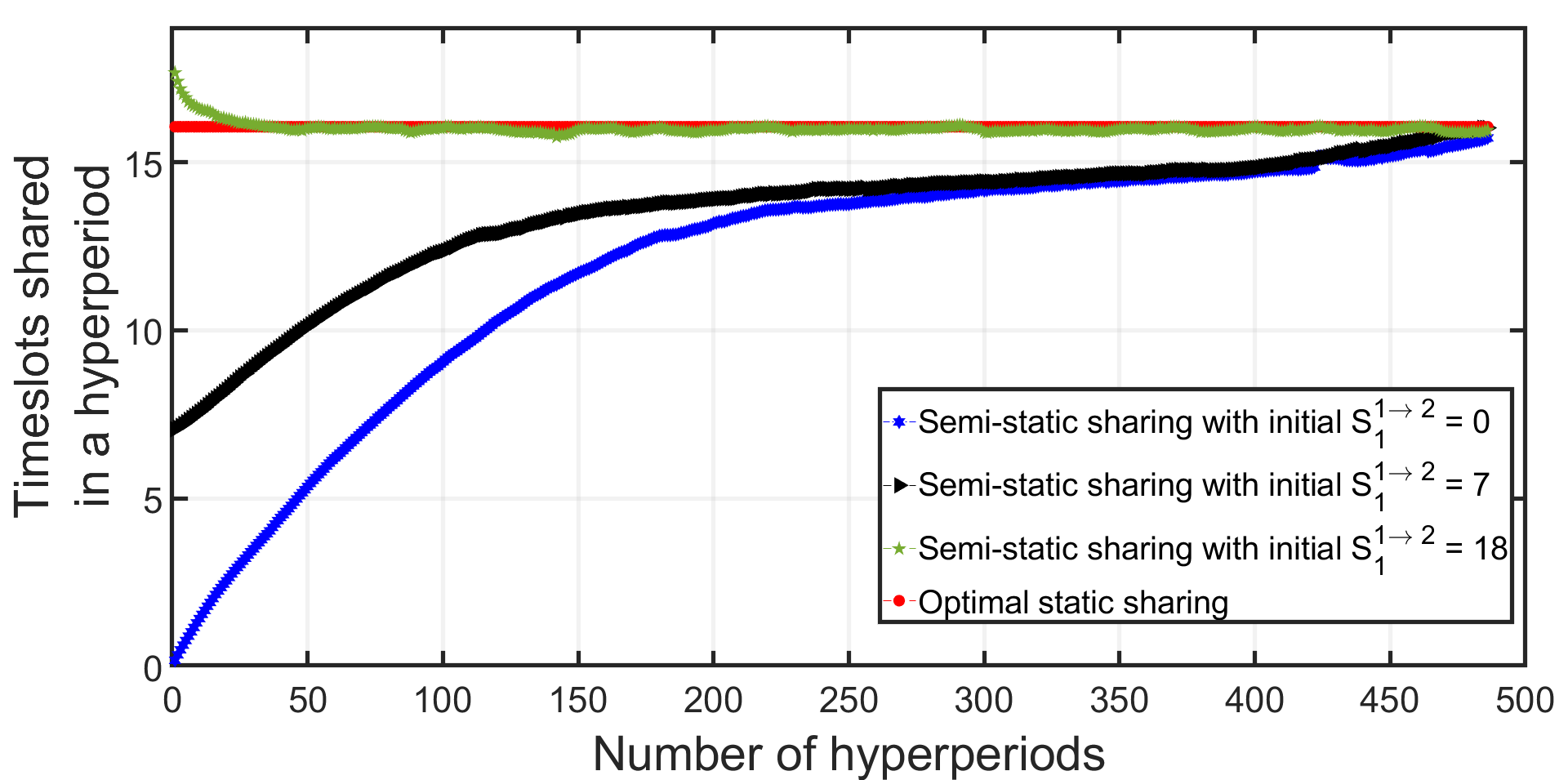}
   \caption{}
   \label{convergence S}
\end{subfigure}
\caption[Iterative semi-static]{Iterative semi-static sharing policy converges to optimal static sharing policy even when initial sharing configurations differ, showing (a) the total QoE, (b) number of timeslots shared in a hyperperiod \label{convergence_diff_S} 
}\end{figure}

\begin{figure}[htbp]
\centering
\begin{subfigure}[b]{0.55\textwidth}
   \includegraphics[width=0.88\linewidth]{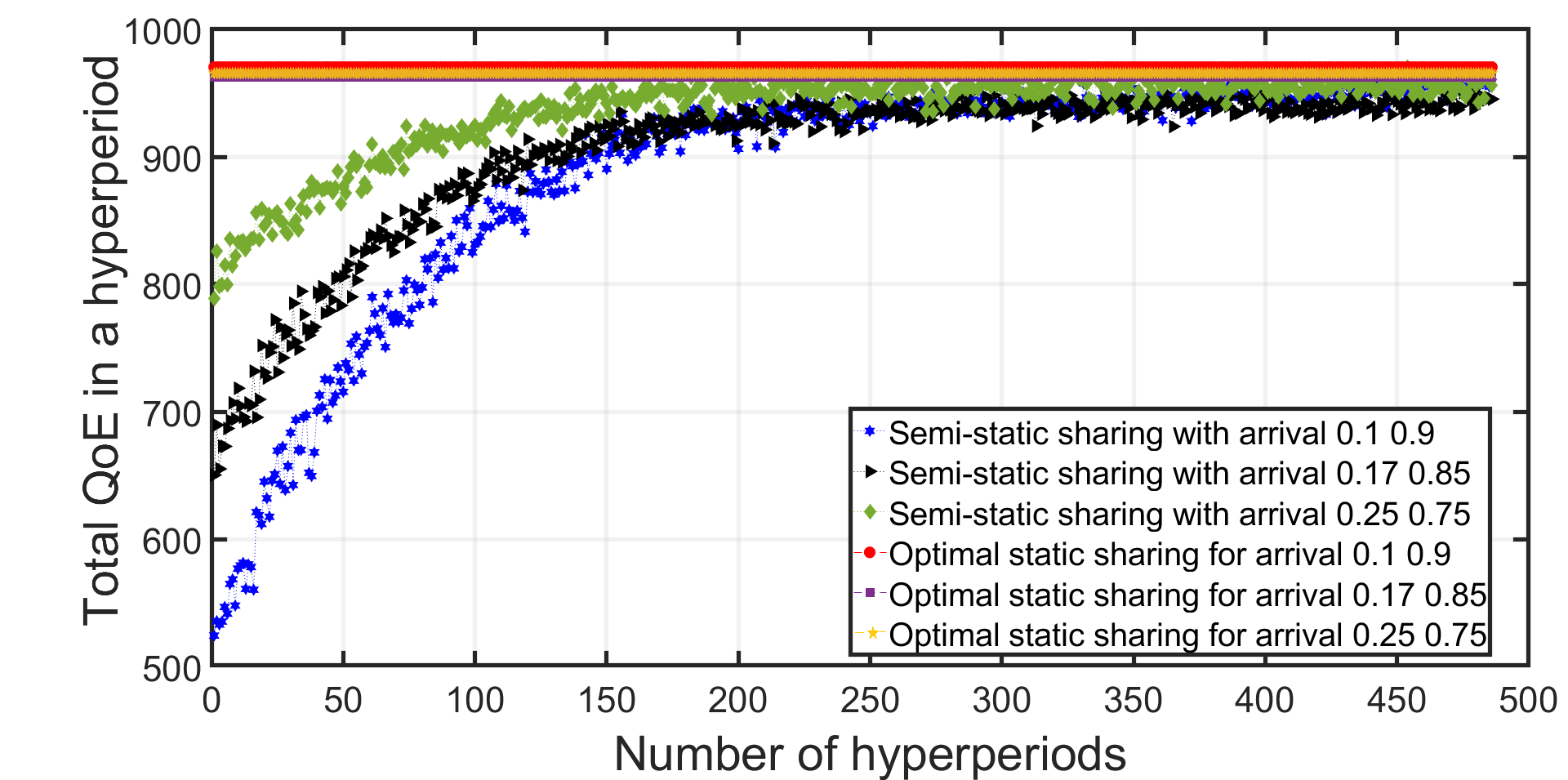}
   \caption{}
   \label{convergence obj imbalance}
\end{subfigure}
\begin{subfigure}[b]{0.55\textwidth}
   \includegraphics[width=0.88\linewidth]{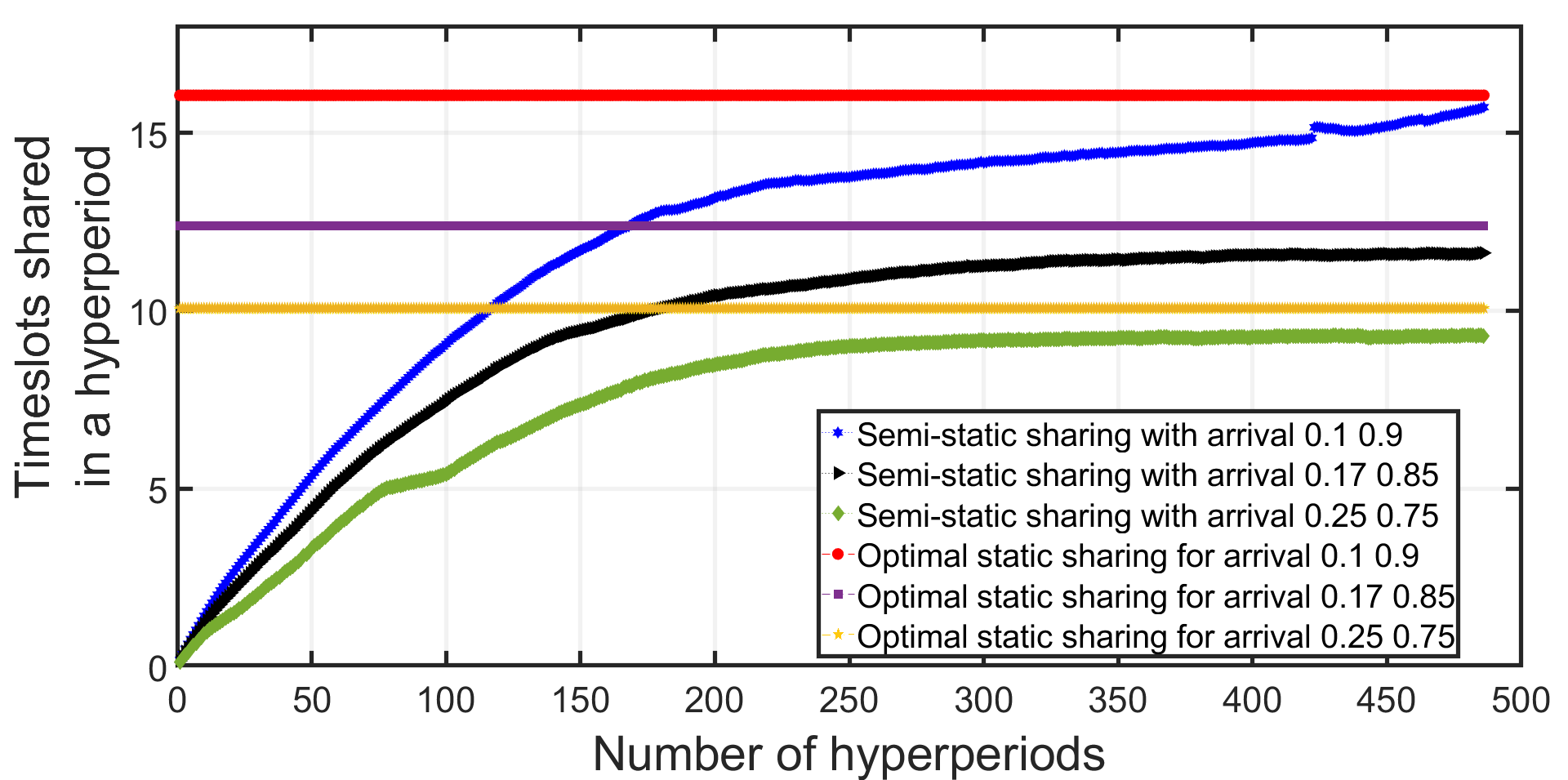}
   \caption{}
   \label{convergence S imbalance}
\end{subfigure}
\caption[Iterative semi-static]{Iterative semi-static sharing policy converges to optimal static sharing policy even when arrival rates differ, depicting (a) the total QoE, (b) number of timeslots shared in a hyperperiod \label{convergence_imb} 
}\end{figure}

From Fig \ref{convergence_diff_S} and Fig \ref{convergence_imb}, we observe that the behavior of iterative semi-static sharing policy converges to that of the optimal static sharing policy. After almost 150 hyperperiods, the iterative semi-static sharing policy is able to achieve objective function close to that obtained by optimal static sharing policy. Even in scenarios where the initial sharing variables differ (Fig \ref{convergence_diff_S}) or the arrival rates vary (Fig \ref{convergence_imb}), the proposed policy consistently converges to the optimal static sharing solution. 

\begin{figure}[htbp]
\centerline{\includegraphics[scale=0.17]{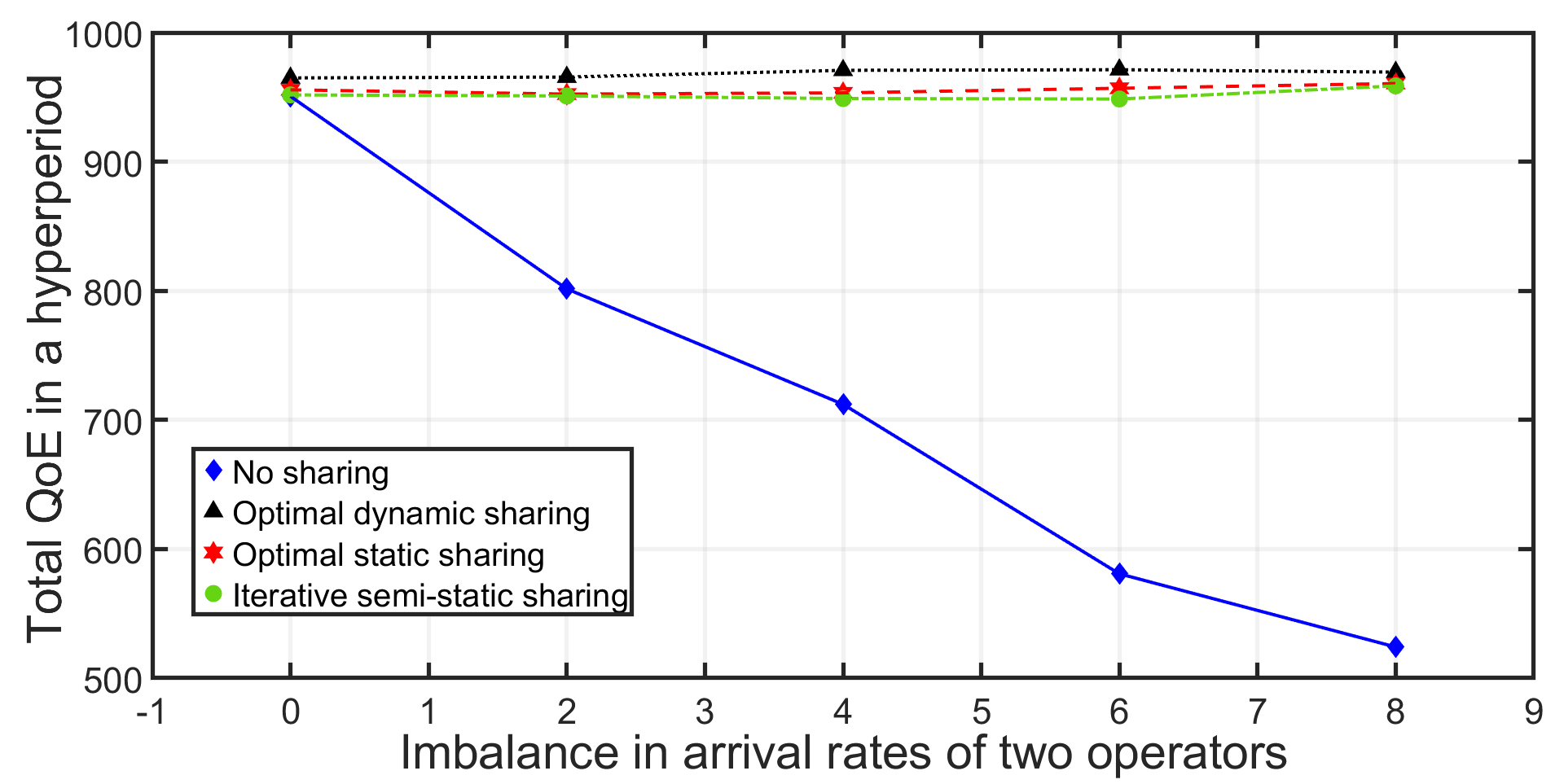}}
\caption{Comparison between different bandwidth-sharing policies}
 \label{comparison} 
\end{figure}

Fig. \ref{comparison} compares the performance of different bandwidth sharing policies w.r.t the no sharing case and provides several important insights. First, it shows that the performance of the optimal static sharing policy is comparable to that of the optimal dynamic sharing policy. Additionally, as the arrival rate imbalance decreases, there is a noticeable drop in percentage improvement of the optimal dynamic sharing policy compared to the no-sharing condition, emphasizing the diminishing gains of dynamic sharing under balanced conditions. Finally, the convergence of the iterative semi-static sharing policy to the optimal static sharing policy is evident, further reinforcing the effectiveness of the static approach in scenarios where iterative adjustments are feasible.


\begin{figure}
\centering
\begin{subfigure}[b]{0.55\textwidth}
   \includegraphics[width=0.89\linewidth]{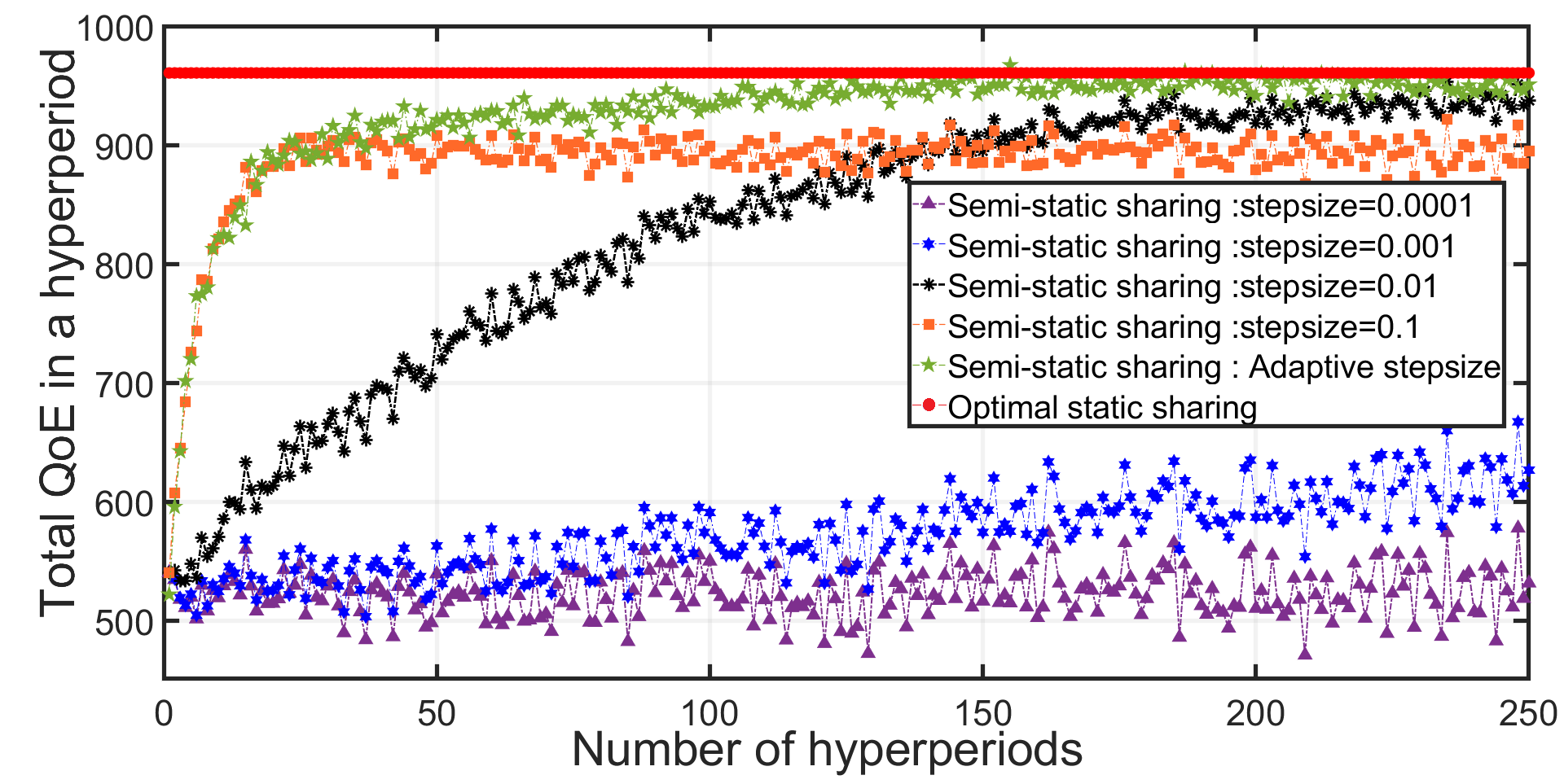}
   \caption{}
   \label{stepsize_obj} 
\end{subfigure}
\begin{subfigure}[b]{0.55\textwidth}
   \includegraphics[width=0.89\linewidth]{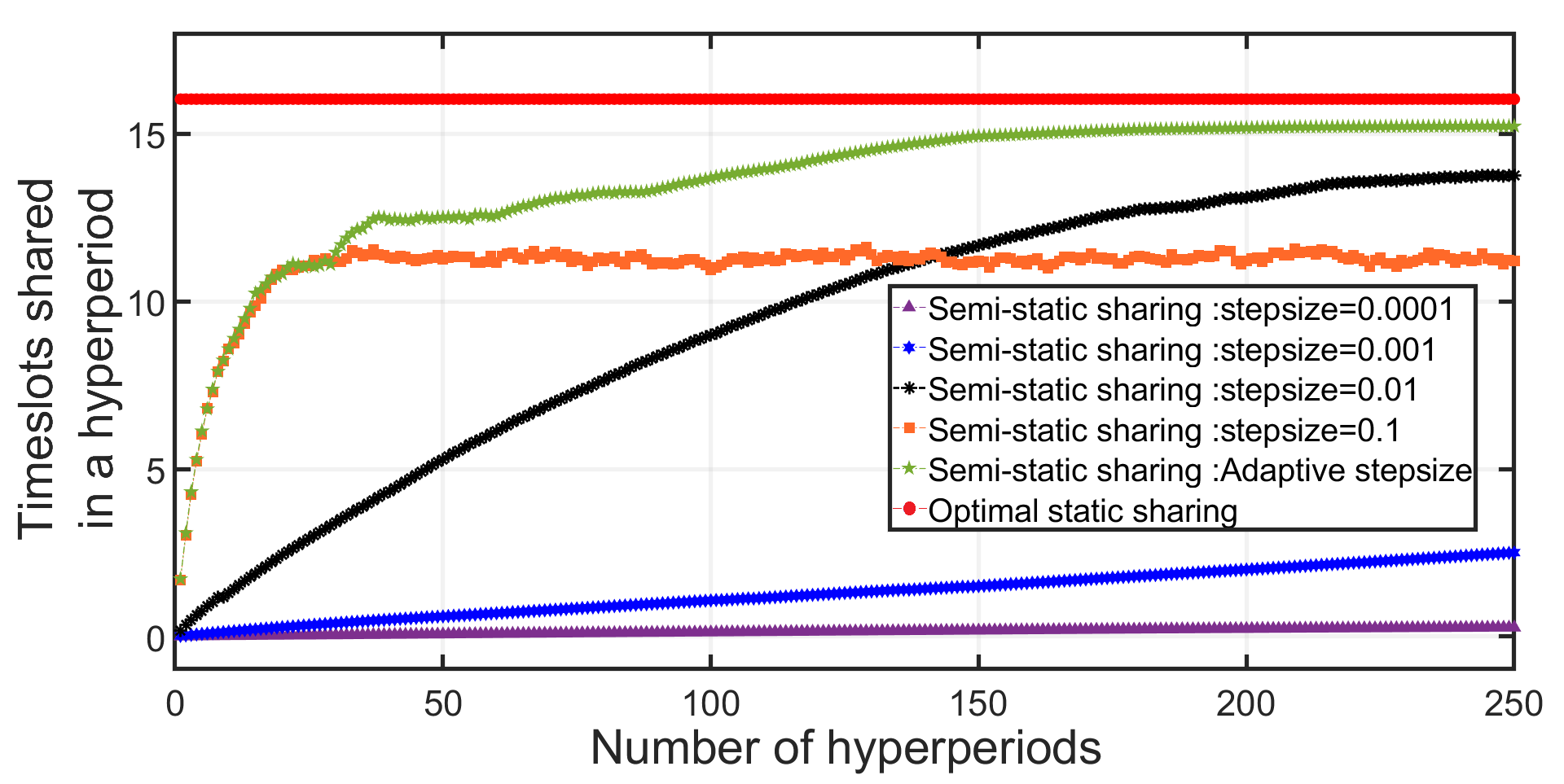}
   \caption{}
   \label{stepsize_S}
\end{subfigure}
\caption[varystepsize]{Effect of different step size values (with a fixed number of iterations) on: (a) the objective function attained, (b) the number of timeslots shared \label{vary step size} 
}\end{figure}


Fig. \ref{vary step size} highlights the impact of varying step sizes on the convergence of the semi-static sharing policy, offering key observations. When a large step size of 0.1 is used, the policy converges quickly but falls significantly short of the optimal value, demonstrating the trade-off between speed and accuracy. Conversely, with a very small step size of 0.0001, the policy achieves higher accuracy but takes an excessively long time to converge. A moderate step size of 0.01 strikes a balance, showing a gradual yet steady increase in performance and converging close to the optimal value within a reasonable number of iterations. Notably, the benefit of using an adaptive or variable step size is evident in the green curve. Starting with a larger step size and progressively decreasing it over time, this approach combines the advantages of rapid initial convergence and precise, stable final performance.



\begin{figure}[htbp]
\centerline{\includegraphics[scale=0.18]{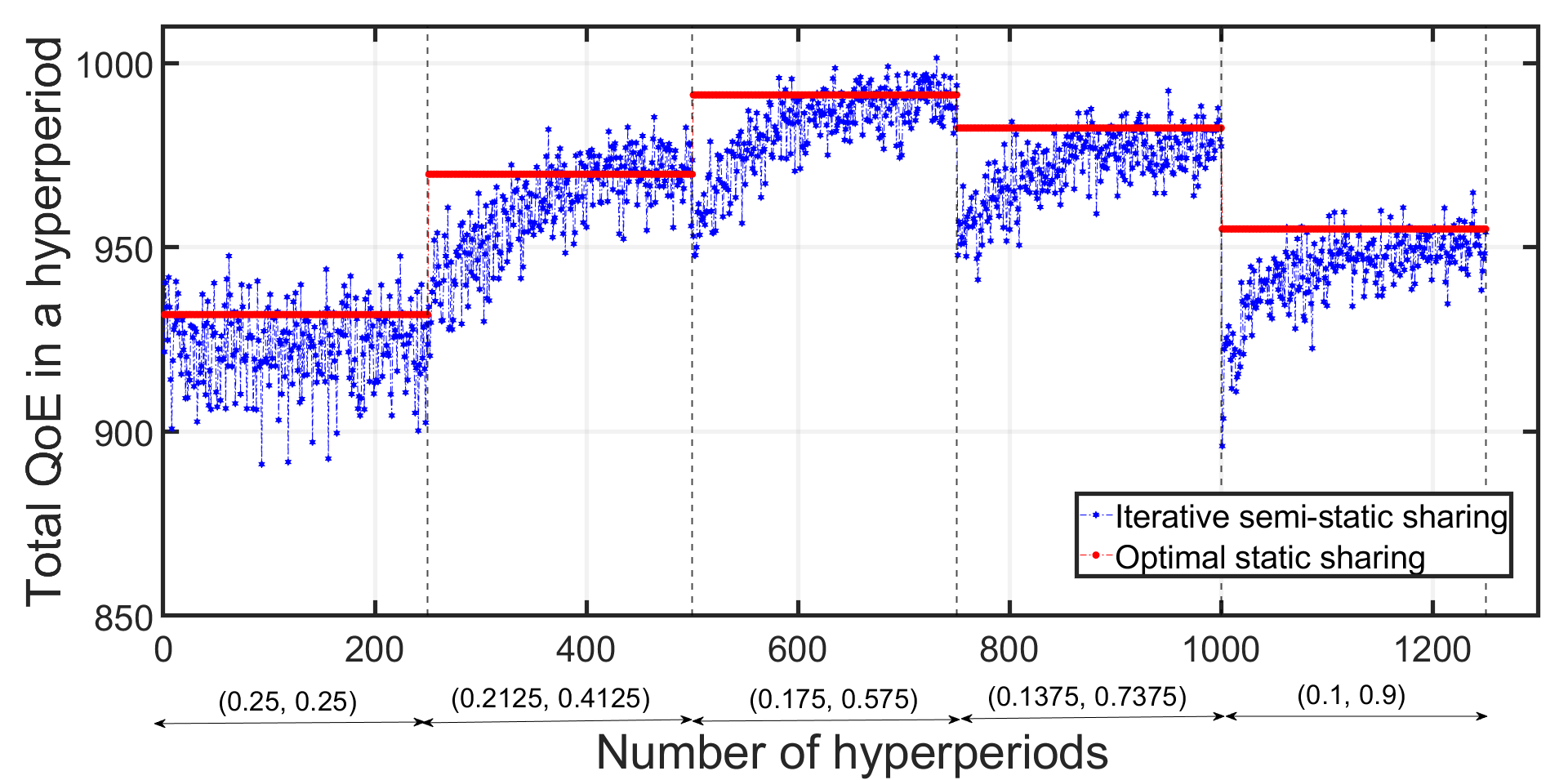}}
\caption{Semi-static sharing policy adapting to changes in the arrival process}
\label{adapting to arrival} 
\end{figure}

\begin{figure}[htbp]
\centerline{\includegraphics[scale=0.18]{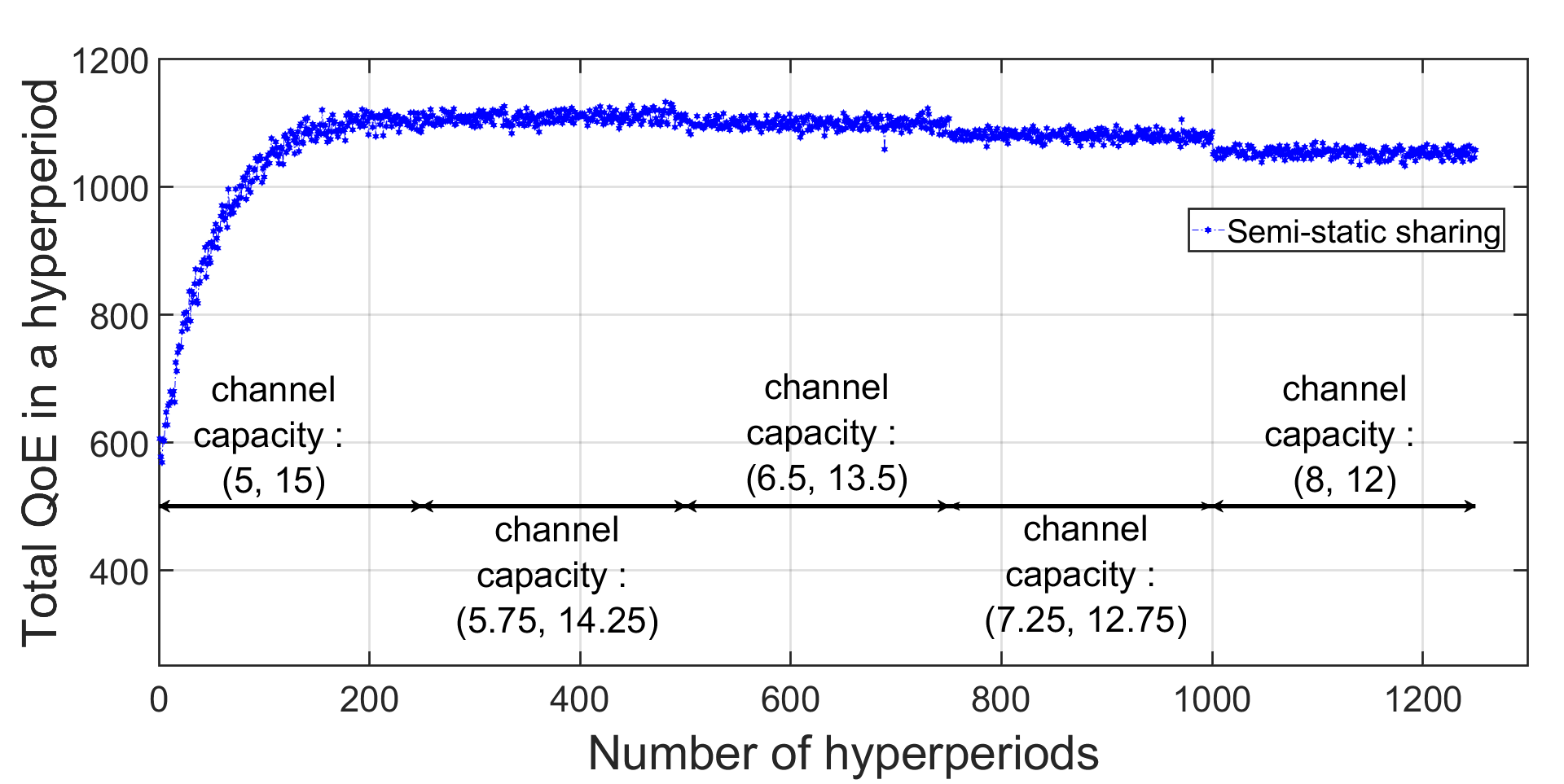}}
\caption{Semi-static sharing policy adapting to changes in the channel capacity}
\label{adapting to C} 
\end{figure}

Fig. \ref{adapting to arrival} and Fig. \ref{adapting to C} demonstrates the adaptability of the semi-static policy in response to changes in the arrival process and variation in channel capacity of the clients respectively, showcasing its effectiveness in handling dynamic conditions. This adaptability allows the policy to maintain efficiency and relevance, even when the arrival patterns fluctuate and channel conditions vary, highlighting its robustness in varying scenarios.

\section{Conclusion} \label{sectionconclusion}
This work demonstrates the effectiveness of our iterative semi-static sharing policy, which consistently converges to the optimal static sharing policy, regardless of initial conditions or arrival rate variations. Our comparative analysis shows that optimal static sharing performs similarly to optimal dynamic sharing, with diminishing benefits of dynamic sharing under balanced conditions.

Additionally, our results highlight the impact of step size and hyperperiod length on policy performance.The semi-static policy effectively adapts to fluctuating arrivals, ensuring robustness across varying scenarios.

A promising direction for future work is the integration of incentive mechanisms to encourage operator cooperation in bandwidth sharing. Since operators may have conflicting interests, designing reward-based or cost-sharing schemes could align incentives and improve participation. 
The framework and performance results in this work act as useful reference points for such extensions. 

\result

\ifCLASSOPTIONcaptionsoff
  \newpage
\fi

\bibliographystyle{IEEEtran}
\bibliography{references2.bib}

\end{document}